\documentclass{article} 
\usepackage[dvipsnames]{xcolor}
\usepackage{iclr2025_conference,times}

\usepackage{amsmath,amsfonts,bm}

\def\eqref#1{equation~\ref{#1}}

\def\1{\bm{1}}

\DeclareMathAlphabet{\mathsfit}{\encodingdefault}{\sfdefault}{m}{sl}
\SetMathAlphabet{\mathsfit}{bold}{\encodingdefault}{\sfdefault}{bx}{n}

\usepackage{url}
\usepackage{graphicx}
\usepackage{listings}
\usepackage{subcaption}
\usepackage{bussproofs}
\usepackage[title]{appendix}

\usepackage{acronym}
\acrodef{pbe}[PBE]{Programming-by-Example}
\acrodef{io}[I/O]{Input-Output}
\acrodef{dsl}[DSL]{Domain-Specific Language}
\acrodef{ai}[AI]{Artificial Intelligence}
\acrodef{agi}[AGI]{Artificial General Intelligence}
\acrodef{arc}[ARC-AGI]{Abstraction and Reasoning Challenge}
\acrodef{regesm}[REGISM]{Repeated Execution-Guided Invoking of Synthesis Model}
\usepackage{tcolorbox}
\usepackage{xcolor}

\title{Shedding Light on Task Decomposition in Program Synthesis: The Driving Force of the Synthesizer Model}

\author{Janis Zenkner, Tobias Sesterhenn, Christian Bartelt\\
    Machine Learning and Cognitive Software\\
    Institute for Software and Systems Engineering, TU Clausthal\\
    \texttt{janis.zenkner@tu-clausthal.de}
    }

\iclrfinalcopy 
\begin{document}

\maketitle
\begin{abstract}
Task decomposition is a fundamental mechanism in program synthesis, enabling complex problems to be broken down into manageable subtasks.
ExeDec, a state-of-the-art program synthesis framework, employs this approach by combining a Subgoal Model for decomposition and a Synthesizer Model for program generation to facilitate compositional generalization. 
In this work, we develop REGISM, an adaptation of ExeDec that removes decomposition guidance and relies solely on iterative execution-driven synthesis.
By comparing these two exemplary approaches—ExeDec, which leverages task decomposition, and REGISM, which does not—we investigate the interplay between task decomposition and program generation.
Our findings indicate that ExeDec exhibits significant advantages in length generalization and concept composition tasks, likely due to its explicit decomposition strategies. 
At the same time, REGISM frequently matches or surpasses ExeDec’s performance across various scenarios, with its solutions often aligning more closely with ground truth decompositions.
These observations highlight the importance of repeated execution-guided synthesis in driving task-solving performance, even within frameworks that incorporate explicit decomposition strategies.
Our analysis suggests that task decomposition approaches like ExeDec hold significant potential for advancing program synthesis, though further work is needed to clarify when and why these strategies are most effective.

\end{abstract}

\section{Introduction}
In 2019, Fran\c{c}ois Chollet introduced the \ac{arc} as a benchmark to advance \ac{agi} research by evaluating systems on their ability to perform broad reasoning tasks~\citep{chollet2019measure}.
Among the many approaches to achieving \ac{agi}, program synthesis has emerged as a particularly promising avenue~\citep{chollet2019measure}. 
Program synthesis automates the creation of programs that meet specified requirements, such as \ac{io} examples~\citep{devlin2017robustfill}, formal logical descriptions~\citep{hocquette2023relational}, natural language instructions~\citep{desai2016program}, or partial program templates~\citep{alur2013syntax}.
This versatility enables systems to focus on the underlying logic of a problem rather than being tied to its specific details and realization. 

Despite recent progress, current systems still fall short of human-level performance~\citep{chollet2024arc}, making human cognition a key inspiration for generalizable reasoning methods.
One such capability is the ability to decompose unfamiliar problems into smaller, manageable components~\citep{marcus2003algebraic, frankland2020concepts}.
Task decomposition leverages this principle by dividing complex tasks into subtasks that require generalizable skills to solve~\citep{gulwani2017program}. 
By solving these subtasks independently and integrating their solutions, systems can tackle complex problems with greater scalability and efficiency.
Moreover, the emphasis on developing general-purpose capabilities, rather than narrowly tailored solutions, aligns with the broader goal of fostering universal reasoning abilities~\citep{chollet2019measure}.

ExeDec~\citep{shi2023exedec} incorporates these two ideas by combining program synthesis with task decomposition. 
As such, ExeDec can be seen as a multi-level transductive-inductive reasoning approach, which has shown to be promising for boosting program synthesis performance~\citep{li2024combining}.
It employs two interconnected components: a \textit{Subgoal} Model that iteratively divides complex tasks into subtasks by predicting the subgoal of the next step (high-level transductive reasoning) and a \textit{synthesis} model that generates programs to solve these subtasks (low-level inductive reasoning). 
This structured framework enables the Subgoal Model to acquire broadly applicable skills essential for addressing diverse and complex tasks, while the synthesis model focuses on their implementation, outperforming large Language Models on two program synthesis domains~\cite{shi2023exedec}.

ExeDec compares its performance against a No-Subgoal Ablation, i.e., a baseline where the Subgoal Model is excluded.
Besides the difference in architecture, the key distinction lies in the training of the synthesis model: In ExeDec, the synthesis model focuses purely on program generation, while in the No-Subgoal Ablation, it is tasked with both implicit task decomposition \textit{and} program generation.
To distinguish both synthesis models, we refer to ExeDec's synthesis model as Synthesizer Model.

\begin{figure}[t]
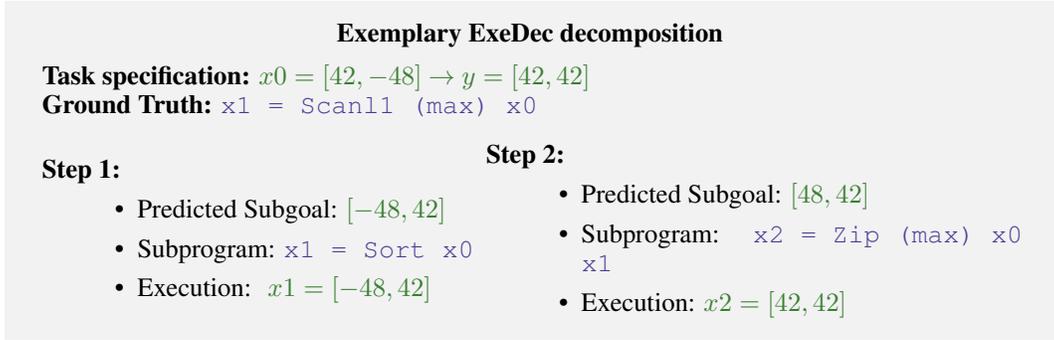

    \begin{tcolorbox}[colback=gray!10, colframe=gray!80, arc=0mm, boxrule=0pt, width=\textwidth]
    \begin{minipage}{\textwidth}
        \begin{center}
            \textbf{Exemplary ExeDec decomposition}
        \end{center}
        \textbf{Task specification:} {\color{OliveGreen}{$x0 = [42, -48]$ $\rightarrow$ $y = [42, 42]$}}\\ \textbf{Ground Truth:} \color{BlueViolet}{\texttt{x1 = Scanl1 (max) x0}}\\
    \end{minipage}
    \begin{minipage}{0.45\textwidth}
        \textbf{Step 1:}
        \begin{itemize}
            \item Predicted Subgoal:  {\color{OliveGreen}{$[-48, 42]$}}
            \item Subprogram: {\color{BlueViolet}{\texttt{x1 = Sort x0}}}
            \item Execution: {\color{OliveGreen}{ $x1 = [-48, 42]$}}
        \end{itemize}
    \end{minipage}
    \begin{minipage}{0.55\textwidth}
        \textbf{Step 2:}
        \begin{itemize}
            \item Predicted Subgoal:  {\color{OliveGreen}{$[48, 42]$}}
            \item Subprogram: {\color{BlueViolet}{\texttt{x2 = Zip (max) x0 x1}}}
            \item Execution:  {\color{OliveGreen}{$x2 = [42, 42]$}}
        \end{itemize}
    \end{minipage}
    \end{tcolorbox}
    \caption{ExeDec workflow illustrating the challenges posed by misleading subtasks. The task is taken from the original test set and can be solved by computing the cumulative maximum of the input list. The subgoals are those generated by ExeDec. The decomposition process is shown for a single \ac{io} pair, while the full workflow with all \ac{io} pairs is available in Appendix~\ref{app:examples}.}
    \label{fig:example}
\end{figure}

A deeper analysis of ExeDec’s underlying mechanisms offers insights into refining program synthesis approaches and optimizing task-solving strategies. 
Our analysis is motivated by the hypothesis that the Subgoal Model plays a nuanced role in boosting ExeDec’s performance and that the repeated invocation of the Synthesizer Model is the major driving factor behind task completion, rather than the Subgoal Model itself.
Figure~\ref{fig:example} illustrates this dynamic with a single-step task, i.e., a task solvable by a single function and without decomposition.
Here, the Subgoal Model generates a subtask prediction that differs from the ground truth ($[42, 42]$) causing the Synthesizer Model to produce a subprogram deviating from the target program.
In the next step, despite another Subgoal Model inaccuracy, the Synthesizer Model corrects the error and ultimately solves the task. 
This highlights the Synthesizer Model’s robustness in mitigating decomposition inaccuracies.
A key implication is that decomposing tasks more frequently—regardless of decomposition accuracy—can be beneficial, as it allows the Subgoal Model to leverage the Synthesizer Model’s corrective capabilities. 
The workflow in Figure~\ref{fig:example} is a real example generated by ExeDec; additional examples are in Appendix~\ref{app:examples}.
The results of the ExeDec study quantitatively support this hypothesis further:
Even though, ExeDec achieves a small boost in performance beyond the No-Subgoal Ablation, the No-Subgoal Ablation brings the most significant performance improvement~\citep{shi2023exedec}.
This suggests that the explicit decomposition adds value but may not be critical for all tasks and that the Synthesizer Model has a greater influence on overall task success than previously assumed.

Building on these observations, this paper explores the interaction between task decomposition and synthesis within ExeDec. Our key contributions are as follows:
\begin{enumerate}
    \item We provide insights into the challenges of learning task decomposition across domains, showing that learning is easier in domains with fewer decomposition possibilities.
    \item We show that the Synthesizer Model plays a key role in boosting performance, even in the absence of explicit \textit{and} implicit decomposition guidance.
    \item We reveal that the Synthesizer Model, in hindsight, decomposes tasks, even without being trained to do so.
\end{enumerate}

\section{Background}
\subsection{\acl{pbe}}
\ac{pbe} enables task specification without requiring programming knowledge, hardware familiarity, or coding skills~\citep{gulwani2011automating}, making it accessible and versatile across various domains.
%
A \ac{pbe} task is defined through \ac{io} pairs that specify a desired behavior (see Figure~\ref{fig:example}).
Formally, the specification of a task is a set $X = {(I_1, O_1), \ldots, (I_n, O_n)}$, where each pair $(I_i, O_i)$ corresponds to an example of the desired behavior.
To formalize the search space for solutions, \ac{pbe} operates within a \ac{dsl}, which defines the set of possible programs $P$. 
The \ac{dsl} encompasses functions, identifiers, constants, and variables, forming the building blocks for program generation. 
The goal in \ac{pbe} is to identify a program $p \in P$ that satisfies the specification. 
In other words, $p$ must transform each input sample $I_i$ into its respective output $O_i$ for all $i \leq n$. 

\subsection{Compositional Generalization}
Compositional generalization refers to a model’s capacity to systematically and predictably generalize to novel combinations of known components, akin to how humans generalize language and concepts~\citep{wiedemer2024compositional}.
To evaluate this ability in program synthesis, \cite{shi2023exedec} introduced specific task categories.
Each task is designed to examine a distinct aspect of generalization, ensuring a comprehensive and diverse evaluation.

\textbf{Length Generalization} Models are trained on tasks requiring $n$ decomposition steps and tested on tasks with $m > n$ decomposition steps. This scenario examines the ability to handle increased program complexity.

\textbf{Compose Different Concepts} This task evaluates whether the model can integrate distinct functional concepts. \ac{dsl} functions are divided into two concepts. Training tasks involve functions either from Concept A or Concept B, while test tasks require the model to combine both concepts.

\textbf{Switch Concept Order} This scenario explores whether models can generalize to new sequences of operations. Again using a categorization of \ac{dsl} functions into two categories, training tasks always begin with Concept A and end with Concept B. Test tasks reverse this order, starting with Concept B and ending with Concept A.

\textbf{Compose New Operations} The goal here is to determine if a model can integrate a previously isolated function into larger compositions. During training, the function is either used alone or excluded, while testing combines it with other functions. A relevant scenario involves synthesizing code that integrates a newly introduced function into a broader workflow, based solely on isolated training examples.

\textbf{Add Operation Functionality} This task evaluates whether a model can infer and apply new functionalities for an operation. Training tasks limit an operation to specific functionalities (e.g., \texttt{Scanl1} with \texttt{min}), while testing introduces variations (e.g., \texttt{Scanl1} with \texttt{max}, or \texttt{*}).

\section{Methods \& Evaluation}
\subsection{ExeDec}
\begin{figure}[t]
    \centering
    \includegraphics[width=.8\textwidth]{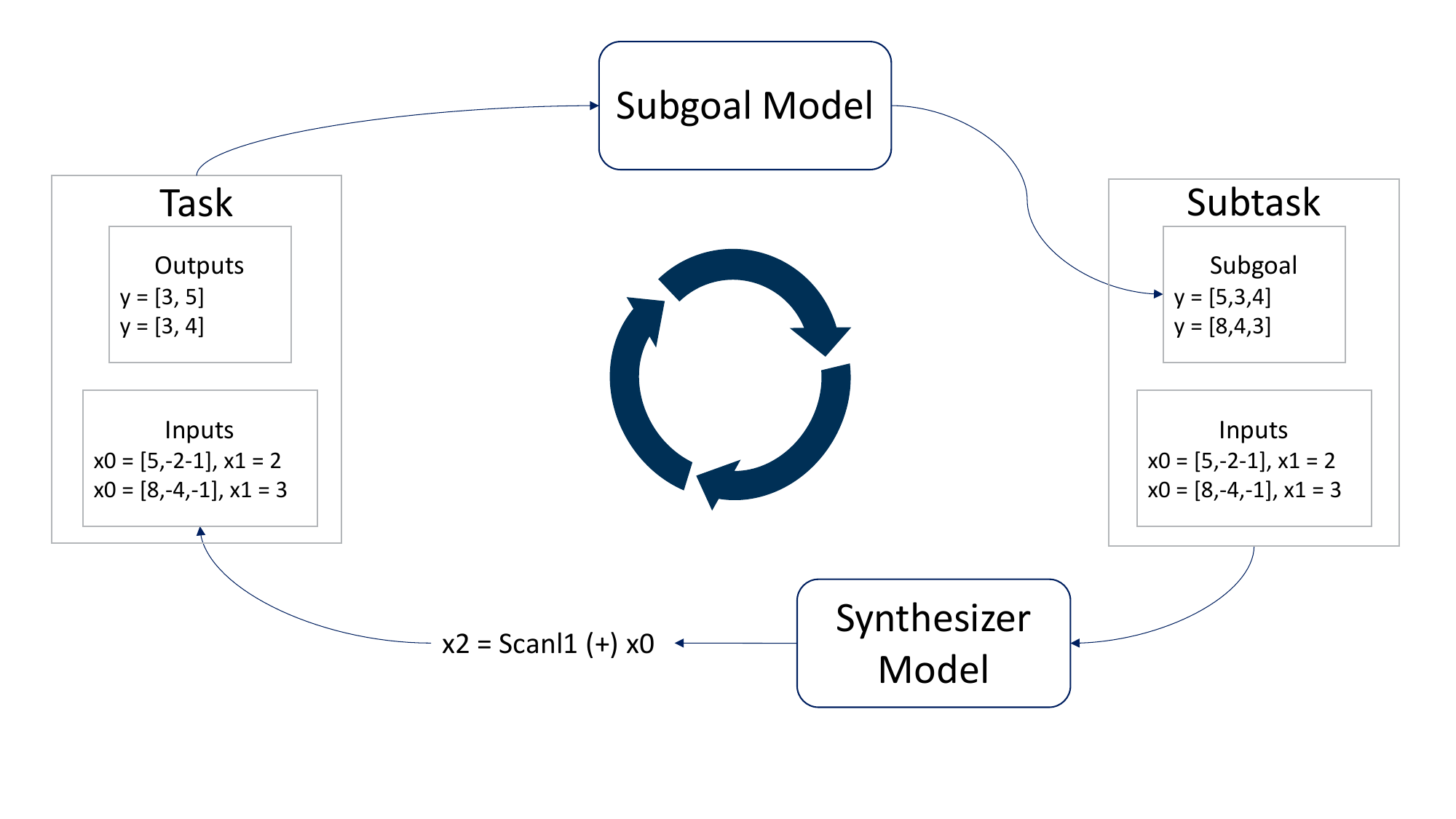}
    \caption{ExeDec Workflow: First, the Subgoal Model predicts the next subgoal based on the current task specifications. Next, the Synthesizer Model generates a program using the subtask specifications, aiming to solve the subtask. Finally, the generated program is executed, and its output is used to update the task specifications.}
    \label{fig:exedec}
\end{figure}

ExeDec~\citep{shi2023exedec} is an approach designed to decompose complex tasks into manageable subtasks and solve them effectively. 
Figure~\ref{fig:exedec} displays the framework which comprises two primary modules: The Subgoal Model, responsible for decomposing the task into subtasks, and the Synthesizer Model for generating subprograms, i.e., programs solving the subtasks.
The generated subprogram is subsequently executed, and its output values are utilized to update the task specification.
Both modules are Transformer models which are trained on decomposed data using a teacher-forcing approach.
For each subprogram in the ground-truth solution, the training data includes:
the updated specification derived from executing the preceding ground-truth subprogram, the execution results of the subprogram on all examples, and the subprogram itself.
The Subgoal Model is trained to predict the subprogram's execution result based on the updated specification.
Given the ground truth \textit{subtask} specifications, the Synthesizer Model is trained to predict the subprogram solving the subtask.
The model used in the No-Subgoal Ablation matches the Synthesizer Model architecture.
Yet, it is trained to predict the subprogram of the next subtask based on the updated \textit{task} specifications. 

\subsection{\acs{regesm}}
To further investigate the role of ExeDec’s Subgoal Model, we developed \ac{regesm}, a modified version of the ExeDec framework that excludes the Subgoal Model.
\ac{regesm} invokes the Synthesizer Model to generate a single-step program, i.e., a program consisting of one function only, executes this program, and updates the task specifications using the execution output.
At a high level, \ac{regesm} shares the same architectural setup and workflow as ExeDec’s No-Subgoal Ablation. The key distinction lies in the training paradigm and the role of the synthesis model.

In both \ac{regesm} and the No-Subgoal Ablation, the synthesis model generates a single-step program, executes it, and updates the task specification using the execution results. This iterative process creates new subtask inputs for subsequent steps.
The primary difference between the two settings lies in how the models are trained. The No-Subgoal Ablation model is trained to predict subprograms using the ground truth subtask inputs and the \textit{task} outputs. This forces the model to implicitly infer both decomposition strategies and program generation, making the learning process more entangled. 
In contrast, the Synthesizer Model used in \ac{regesm} is trained to predict subprograms using ground truth subtask inputs and \textit{subtask} outputs, as it is originally designed for use with ExeDec's decomposition model. Consequently, the Synthesizer Model is trained to focus purely on program generation without the added complexity of decomposition.
During testing, when subtask outputs are unknown, both models receive subtask inputs and task outputs as input. This distinction in training objectives enables us to assess the impact of explicitly supervising decomposition strategies, as done in ExeDec, compared to relying purely on program synthesis, as in \ac{regesm}. 

\subsection{Evaluation}
\subsubsection{Domains}
\textbf{String Manipulation}
The RobustFill domain, introduced by \cite{devlin2017robustfill}, focuses on synthesizing sequences of string manipulation operations.
Given \ac{io} string pairs, the goal is to generate programs that transform inputs into corresponding outputs using a \ac{dsl} (Appendix~\ref{app:dsls}). 
The \ac{dsl} includes functions for extracting substrings, modifying strings, and composing operations by applying modifications to prior results. It also supports constant string characters.
RobustFill programs are structured as concatenations of largely independent expressions, except for the \textit{Compose} operation, which applies a function to another’s result.
This weak interdependence among subtasks reduces the combinatorial space, simplifying task decomposition learning. 
However, it also limits error correction during synthesis, making precise decomposition and implementation critical for correct program generation.

\textbf{List Manipulation}
The Deepcoder domain \citep{balog2016deepcoder} addresses tasks involving integer list manipulations.
Its \ac{dsl} includes both first-order and higher-order operations.
Details can be found in Appendix~\ref{app:dsls}.
Tasks in this domain can have multiple inputs, and the outputs are defined as either a list of integer values, an integer, or a boolean value.
Programs follow a line-by-line structure, where each expression relies on an input variable and/or the result of a previous expression as its input.
Such a framework mirrors the way humans typically construct programs.
The domain’s line-by-line style enables error correction yielding an expansive combinatorial space and various task decomposition strategies contributing to its inherent complexity.

\subsubsection{Experimental Setup}
For a fair comparison with ExeDec, \ac{regesm} invokes the Synthesizer Model the same number of times as ExeDec’s Subgoal Model decomposes the task, with all approaches using a beam size of ten.
Furthermore, we used the same 1,000 test tasks, the pretrained models, and \acp{dsl} from the original ExeDec study~\citep{shi2023exedec}\footnote{https://github.com/google-deepmind/exedec}.
Details on the creation of the test tasks can be found in Appendix~\ref{app:benchmarks}.
Five different initalizations were used per task category.
Results are displayed as averages across those initializations, and error bars denote the 95\% confidence interval

In addition to ExeDec and \ac{regesm}, we evaluate the Synthesizer Model itself to assess the effect of step-by-step as opposed to single-step generation.
In summary, we compare ExeDec against these approaches:
First, in the No-Subgoal Ablation setting, the synthesis model is trained to predict the subprogram using the subtask inputs and the task output. This setup requires the model to simultaneously learn both decomposition strategies and general-purpose problem-solving abilities.
Second, in REGISM, the Synthesizer Model is trained to predict the subprogram using the in- and outputs of the subtask. This approach focuses exclusively on enabling the model to generate a program based on the provided specifications.
Third, we evaluate the Synthesizer Model itself by invoking it only once per task.

\section{Results \& Discussion}
Our analysis is motivated by the hypothesis that the Subgoal Model plays a nuanced role in boosting ExeDec's performance. 
Specifically, we hypothesize that the repeated invocation of the Synthesizer Model is the major driving factor behind task completion, rather than the Subgoal Model itself. 

\subsection{Domain Differences and Quality of the Subgoal Model}
\begin{figure}[h!]
    \centering
    \includegraphics[width=0.48\textwidth]{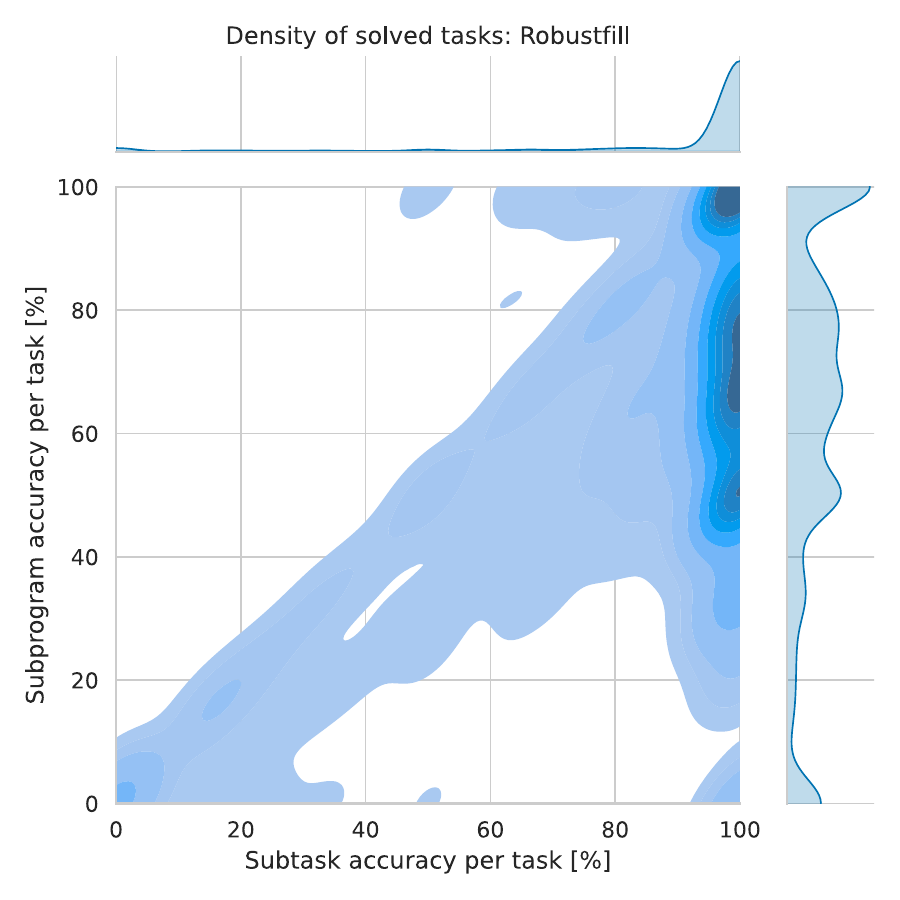}
    \includegraphics[width=0.48\textwidth]{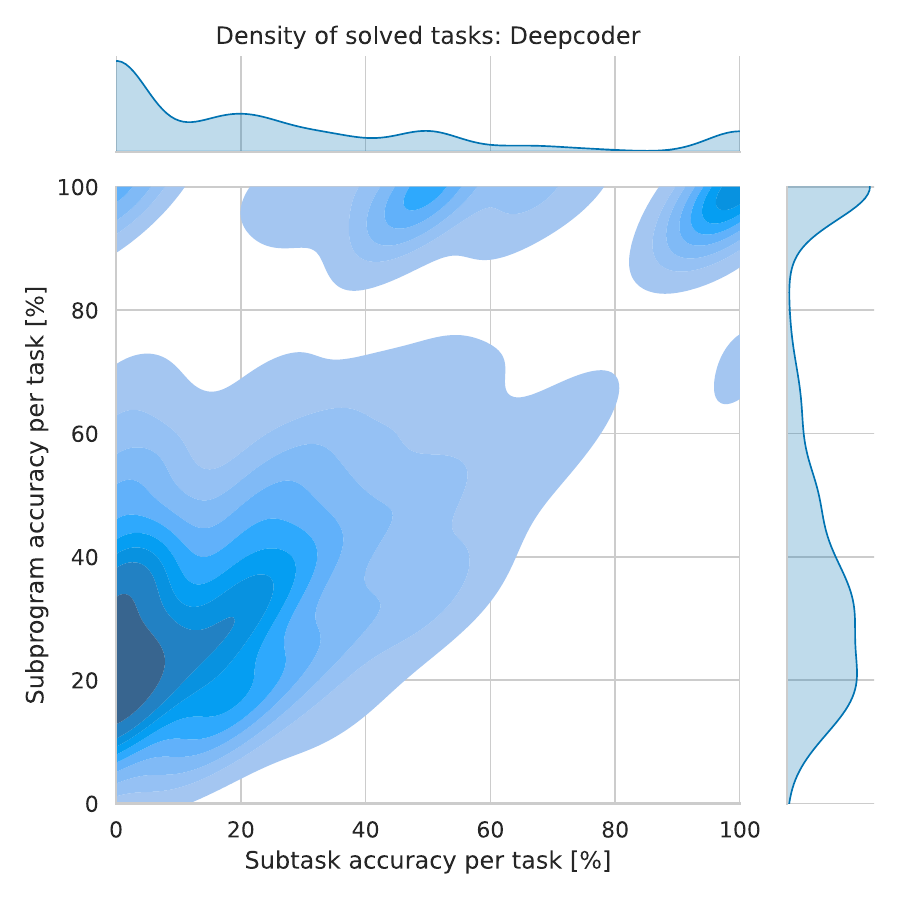}
    \caption{Tasks solved by ExeDec, clustered based on subtask and subprogram accuracy. The x-axis represents subtask accuracy per task (overlap between predicted and ground truth subtasks), while the y-axis indicates subprogram accuracy (overlap between predicted programs and ground truth solutions). Values are averaged across compositional generalization categories, based on 8,900+ solved tasks in DeepCoder and 26,000+ in RobustFill.}
    \label{fig:average_densities_both_domains}
\end{figure}

To evaluate ExeDec’s ability to learn task decompositions, we analyze its performance relative to the ground truth.
Figure~\ref{fig:average_densities_both_domains} presents a density plot illustrating the distribution of solved tasks. 
The x-axis represents the subtask accuracy per task, measured as the overlap between predicted subtasks and ground truth subtasks. 
The y-axis indicates subprogram accuracy, defined as the overlap between predicted programs and the ground truth solutions. 
Each task’s position in the plot reflects its accuracy: for a task requiring four decomposition steps, if ExeDec correctly predicts two subtask specifications and solves three subtasks with ground-truth-matching programs, it appears at~$(x=50\%, y=75\%)$. 
Consequently, the density plot highlights distinct solution patterns.
In the top-right quadrant, both subtasks and subprograms closely align with the ground truth. 
The bottom-right region represents cases where subtasks mostly match the ground truth, but subprograms differ due to alternative syntactic implementations. 
The bottom-left quadrant contains instances where both subtasks and subprograms deviate from the ground truth. 
Finally, in the top-left region, subprograms align closely with the ground truth while subtasks differ. 
This visualization provides insight into the semantic and syntactic accuracy of ExeDec’s solutions.

\textbf{Ambiguous Performance of the Subgoal Model}
In the Robustfill domain, the majority of tasks is clustered in the bottom and top right quadrant, i.e., subtasks are predicted as intended by the ground truth but the implementation differs, showcasing the Subgoal Model's accuracy in learning decompositions.
A smaller portion of task solutions stretches along the main diagonal. 
These tasks are solved using - at least partially - alternative decompositions.
In the Deepcoder domain, a small cluster exists in the top-right corner, indicating tasks solved as intended by the ground truth.
Yet, the majority of tasks is solved by using alternative decompositions and implementations (bottom-left corner).
In the ExeDec study, the authors hypothesize that the low subgoal accuracy is caused by slight deviations in the predicted subgoals that the Synthesizer Model compensates for (top-left quadrant).
However, as most subtasks are also solved using alternative programs, our results indicate a solution strategy deviating from the ground truth.
Although the tasks are ultimately solved, these results question whether the Subgoal Model plays a crucial role in achieving this.
Plots displaying each compositional generalization category separately can be found in Appendix~\ref{app:results}.

\begin{figure}[h!]
    \centering
    \includegraphics[width=.775\textwidth]{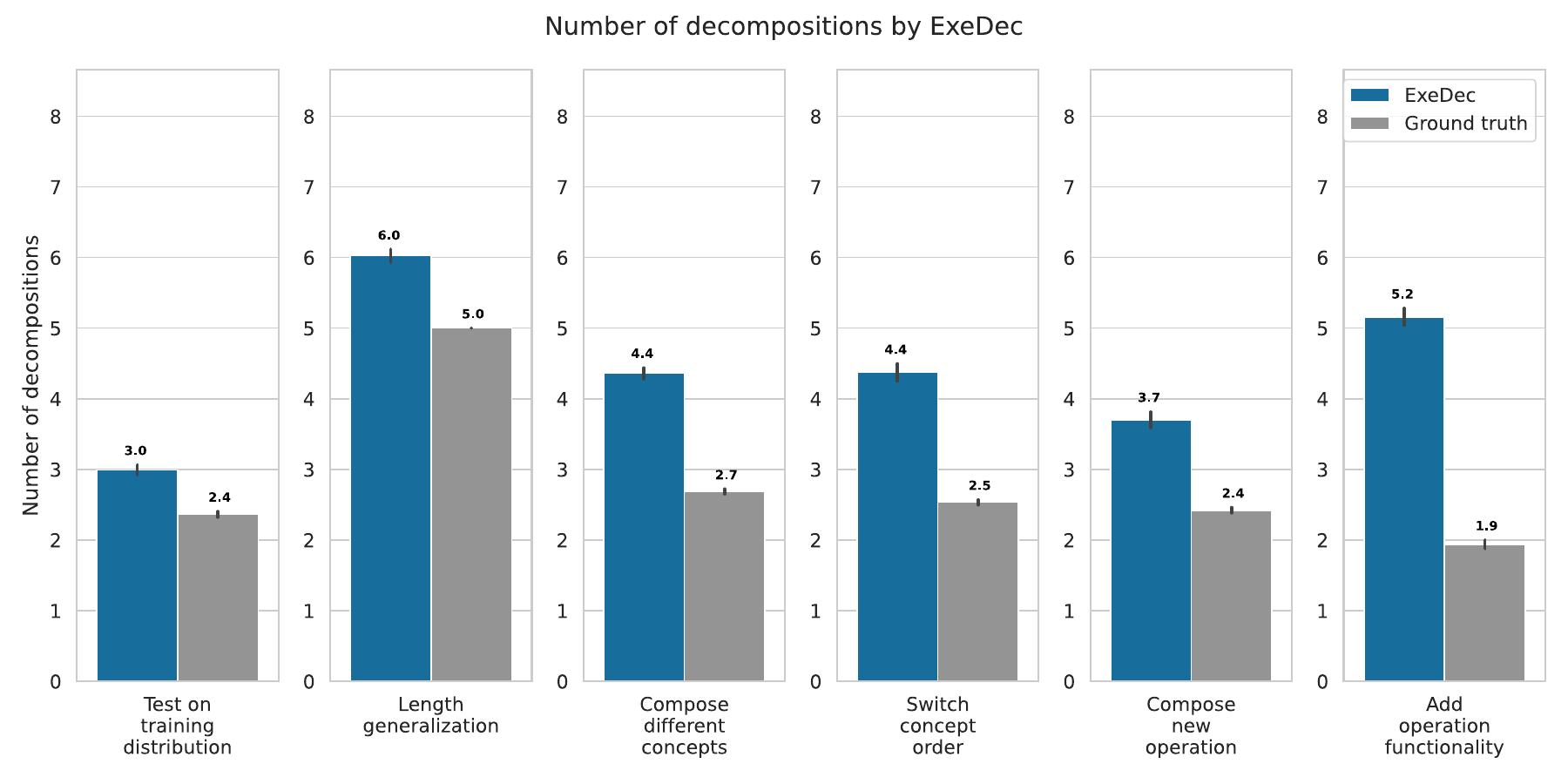}
    \caption{Comparison of the number of decompositions of task solution in the DeepCoder domain using ExeDec.}
    \label{fig:num_decomps_deepcoder_exedec}
\end{figure}

\textbf{More Frequent Invocation of the Synthesizer Model}
The nuanced role of ExeDec's Subgoal Model is further supported by its tendency to decompose tasks into a greater number of subtasks than those specified by the ground truth in the Deepcoder domain~(Figure~\ref{fig:num_decomps_deepcoder_exedec}). 
This increased number of decompositions leads to more frequent invocation of the Synthesizer Model, which seems to be the primary factor driving the observed performance improvements. 

\textbf{Domain differences}
In contrast, in the Robustfill domain, the number of decompositions closely aligns with the ground truth~(Figure~\ref{fig:num_decomps_robustfill}).
This distinct difference can be attributed to the varying difficulty of learning decomposition across the two domains.
Given that the Deepcoder domain better aligns with how humans approach coding and the Subgoal Model performs significantly worse in this domain, we focus on the Deepcoder domain in the subsequent experiments to better understand the reasons behind its performance.

\subsection{REGISM as a Driving Force in Boosting Performance}
\begin{figure}[h]
    \centering
    \includegraphics[width=.8\textwidth]{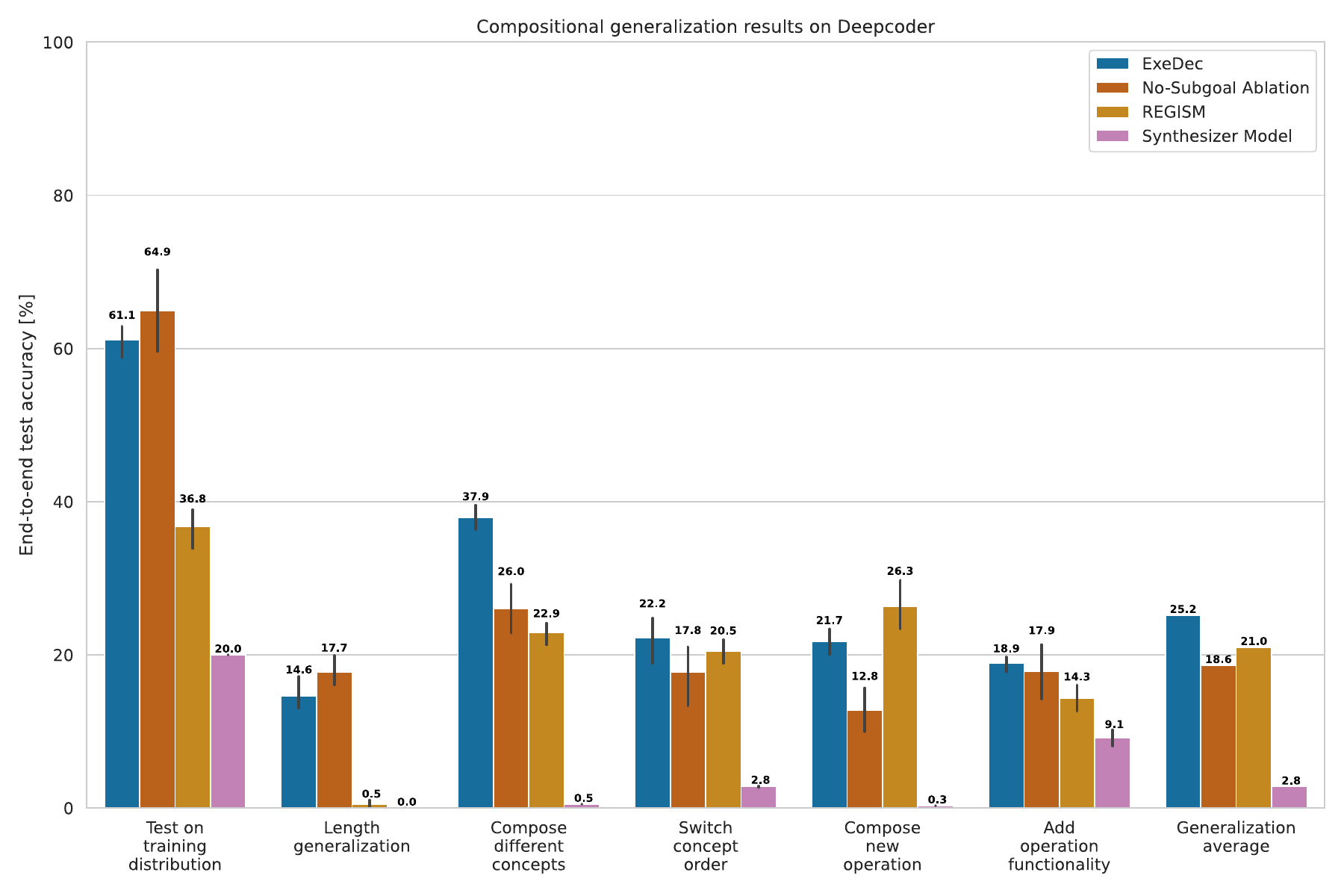}
    \caption{Compositional generalization results using a beam size of 10. End-to-end test accuracy resembles the relative number of solved test tasks.}
    \label{fig:accuracies}
\end{figure}

\textbf{\ac{regesm} performs close to ExeDec on average}
To assess the impact of repeated Synthesizer Model invocation, we evaluate \ac{regesm} on the same tasks as ExeDec. As shown in Figure~\ref{fig:accuracies}, \ac{regesm} outperforms the sole Synthesizer Model, highlighting the benefits of execution guidance and repeated invocation. ExeDec achieves higher accuracy on the training distribution and excels in tasks requiring the composition of different concepts, leveraging decomposition to identify and integrate relevant subcomponents.
\ac{regesm} struggles with length generalization, likely due to its single-step training paradigm, which makes deriving correct subprograms for multi-step tasks more challenging. However, when excluding length generalization tasks, \ac{regesm} performs only slightly worse than ExeDec, suggesting that repeated invocation is a key factor in solving tasks. While ExeDec’s decomposition strategies improve performance in some categories, they offer limited or no advantage in others and can even hinder performance in certain cases. The comparable performance of \ac{regesm} and ExeDec indicates that task-solving effectiveness primarily stems from repeated invocation, independent of explicit or implicit decomposition guidance.

\begin{figure}[t]
    \centering
    \includegraphics[width=.8\textwidth]{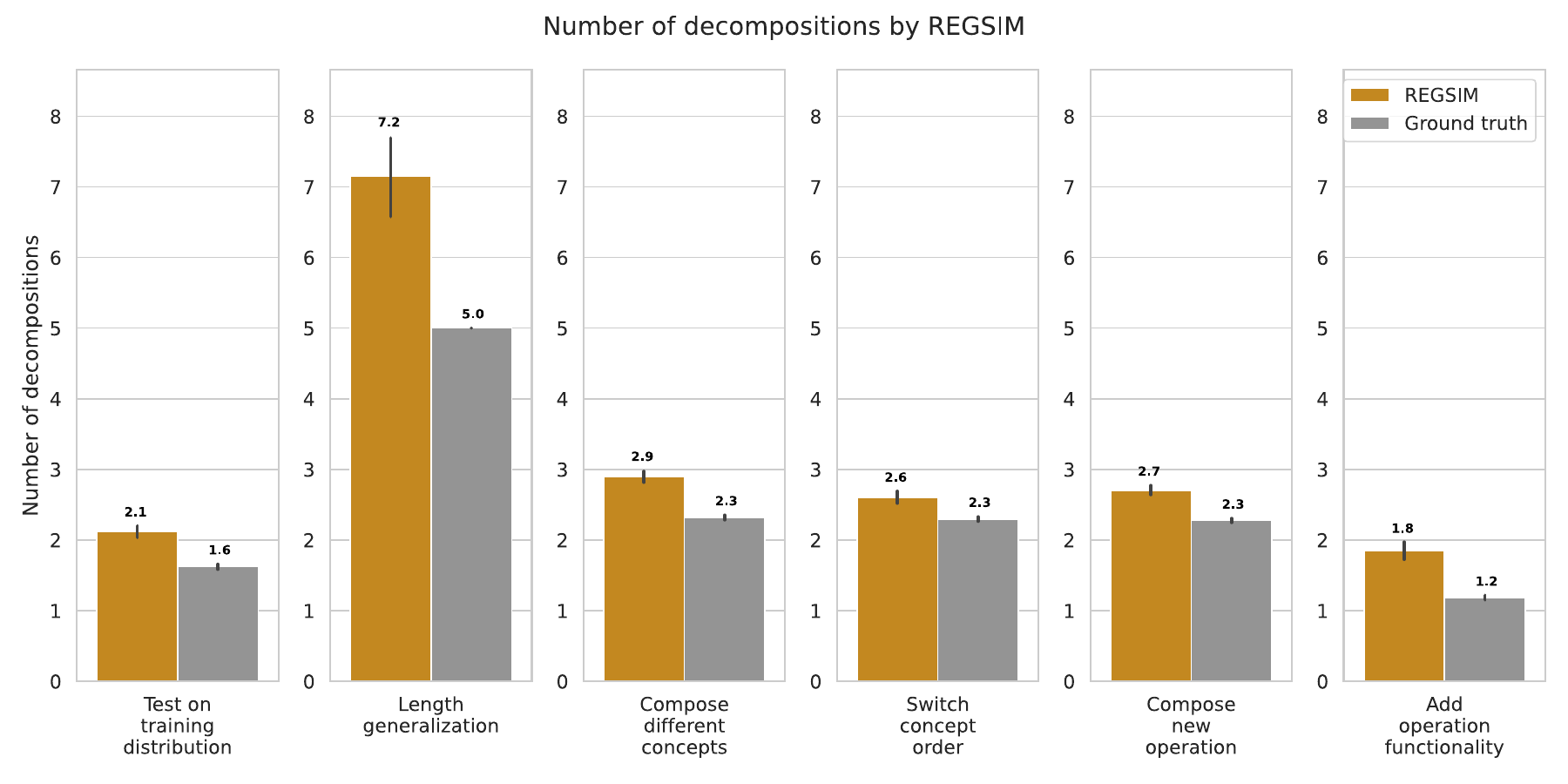}
    \caption{Frequency of invoking the Synthesizer Model in \ac{regesm} relative to the number of ground truth decompositions in the Deepcoder domain.}
    \label{fig:num_decomps_regesm}
\end{figure}

\begin{figure}[h]
    \centering
    \includegraphics[width=0.425\textwidth]{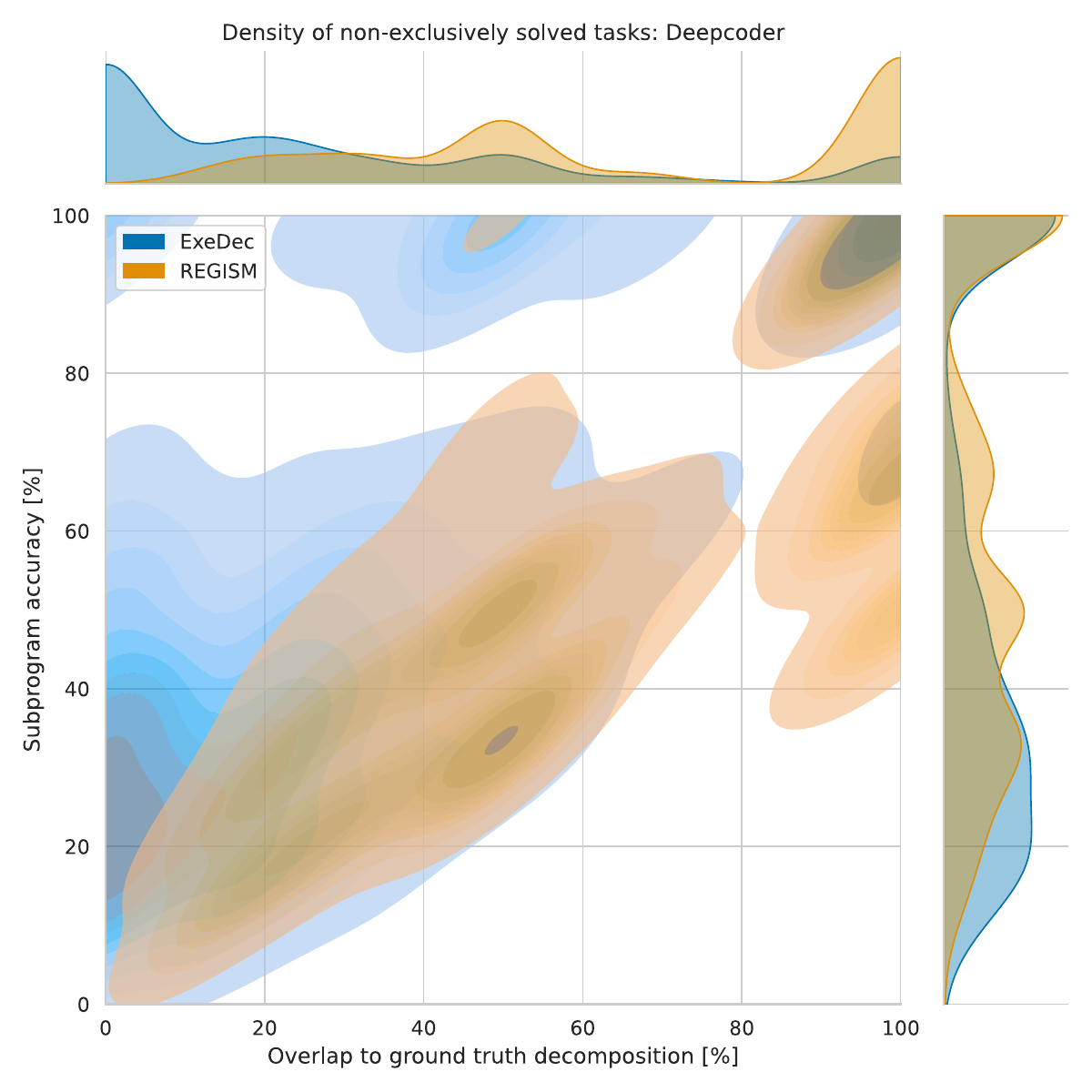}
    \includegraphics[width=0.425\textwidth]{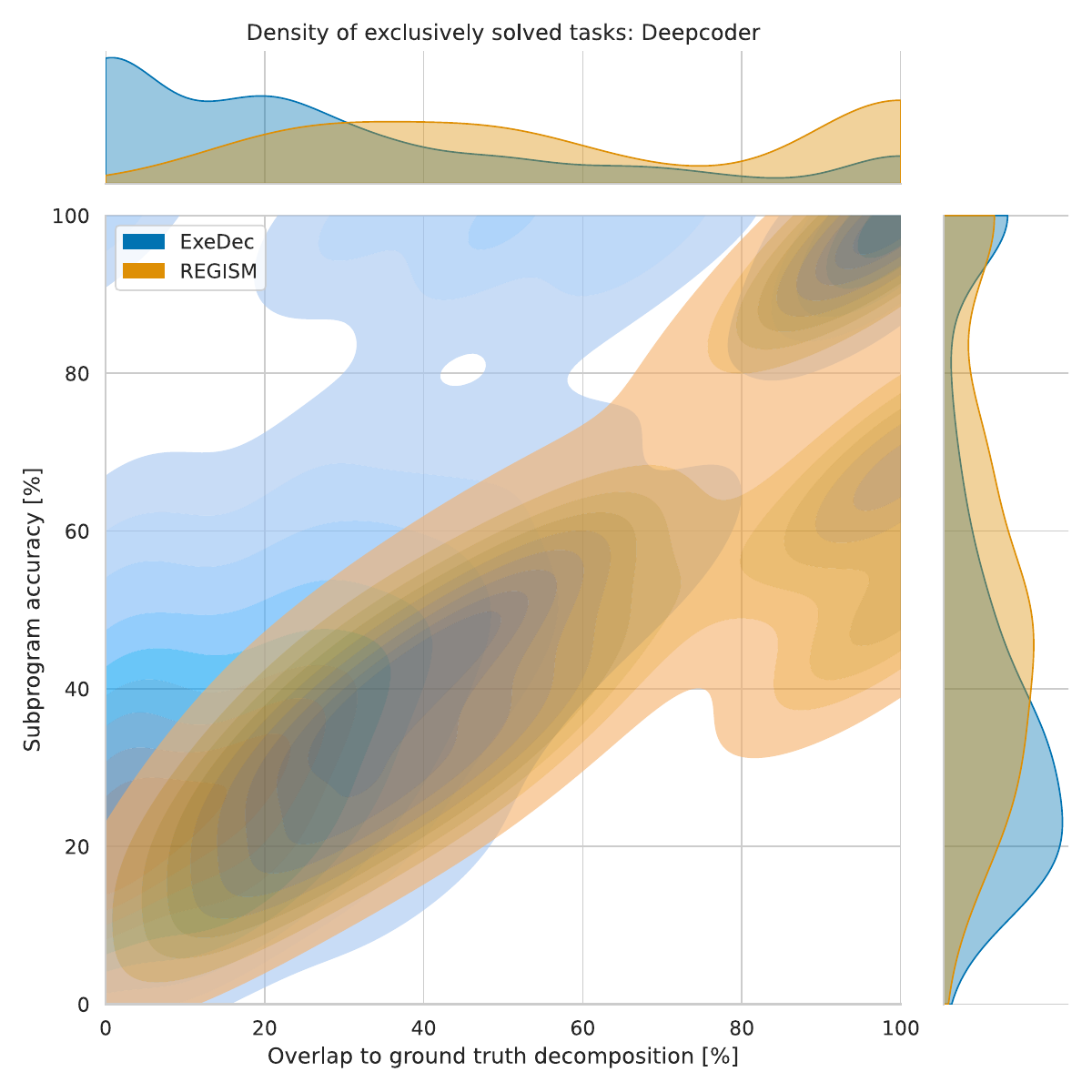}
    \caption{Average densities of ExeDec and \ac{regesm} across all compositional generalization categories and initializations. No clear differences between exclusively and non-exclusively solved tasks can be seen.}
    \label{fig:density_regesm}
\end{figure}

\textbf{\ac{regesm} solves tasks closer to the ground truth}
To gain a better understanding of how \ac{regesm} solves tasks, we again perform a qualitative analysis of the task solutions.
As shown in Figure~\ref{fig:num_decomps_regesm}, \ac{regesm} closely approximates the ground truth number of decompositions.
Comparing \ac{regesm}'s results to ExeDec (Figure~\ref{fig:num_decomps_deepcoder_exedec}), the tasks solved by \ac{regesm} proof to be of similar difficulty to those solved by ExeDec because the number of ground truth decompositions is comparable.
Moreover, \ac{regesm} solves tasks in about as many iterations as decompositions are intended by the ground truth, i.e., in significantly fewer steps than ExeDec.
This indicates that \ac{regesm} solves tasks closer to the ground truth solution thus raising the hypothesis that \ac{regesm}'s inherent decompositions are closer to the ground truth decomposition than the explicit ones of ExeDec. 
To investigate this, we plot the subprogram accuracy against the overlap with the ground truth decomposition after executing the subprograms generated by \ac{regesm} (Figure~\ref{fig:density_regesm}). 
This analysis is further split by tasks solved by both approaches or exclusively by either approach.
Our findings reveal several notable patterns:
\ac{regesm}'s task solutions tend to spread along the main diagonal. 
There is one prominent cluster at the right edge of each plot, where all subtasks align with the ground truth decomposition. 
To a similar but lesser extent than in ExeDec, \ac{regesm}’s largest cluster is located in the bottom-left quadrant.
However, \ac{regesm}’s solutions exhibit a more uniform distribution along the x-axis compared to ExeDec, whose distribution is distinctly shifted toward lower subtask accuracies. 
This effect is slightly more pronounced for tasks solved exclusively by \ac{regesm} than for those solved by both approaches. 
These findings indicate that the Synthesizer Model within \ac{regesm} generates solutions closer to the ground truth than ExeDec despite not being explicitly or implicitly trained for this purpose.
However, for both approaches, the quality of decompositions does not differ significantly between tasks solved exclusively and those solved by both methods.
This finding leaves open the question of why ExeDec solves certain tasks that \ac{regesm} does not, especially when the predicted subgoals in those tasks are not more precise than in tasks solved by \ac{regesm}, and vice versa.

\subsection{Limitations}
While our findings highlight \ac{regesm} as a key component of ExeDec, two limitations warrant consideration. 
First, our evaluation is restricted to a single domain, which limits the generalizability of the results. 
Second, only one decomposition approach was tested, leaving open the question of how alternative methods might influence performance.
\section{Related Work}
\textbf{Programming-by-Example} 
Neural networks play a crucial role in guiding program synthesis for \ac{pbe}~\citep{balog2016deepcoder, yin2017syntactic, lee2018accelerating}.
Many approaches divide synthesis into multiple steps, enabling iterative refinement or task planning.
Search-based methods include top-down approaches that recursively expand partial solutions~\citep{nye2019learning} and bottom-up methods that solve smaller subproblems before combining them into complete programs~\citep{shi2022crossbeam, shi2023lambdabeam, odena2020bustle}.
Planning-based techniques first generate abstract program structures, which are later refined into executable code~\citep{murali2017neural, nye2019learning, chen2020compositional, hong2021latent, klinger2023compositional, prasad2023adapt, zhang2023planning, cano2023learning}. Execution-guided approaches iteratively refine programs using intermediate feedback, such as partial executions or state traces~\citep{ellis2019write, chen2018execution, shrivastava2021learning}.
Recent advances in Large Language Models have further improved program synthesis by leveraging pre-training on large-scale corpora of code and natural language~\citep{li2024programming, li2024combining, tang2024worldcoder}. 
These models enhance synthesis frameworks by capturing semantic and structural patterns of code, aiding in both program generation and refinement.

\textbf{Decomposition-based Program Synthesis} 
Multi-step synthesis methods inherently break tasks into subtasks as part of their iterative process. 
In contrast, decomposition approaches explicitly partition a task \textit{before} leveraging this structure to improve synthesis. 
Some methods model relationships between subprograms, enabling relational reasoning to identify dependencies and shared structures across subtasks~\citep{hocquette2023relational, hocquette2024learning, hocquette2024relational}. 
Similarly, hierarchical composition techniques systematically construct programs by progressively assembling higher-level abstractions from lower-level components~\citep{liu2023hierarchical}.

\section{Conclusion}
Our results highlight ExeDec’s strong performance in the Robustfill domain, where learning decompositions is more straightforward.
At the same time, we demonstrate that the Synthesizer Model plays a critical role in enhancing program synthesis performance in the Deepcoder domain, even in the absence of both explicit and implicit decomposition guidance.
Our findings further reveal that the Synthesizer Model inherently decomposes tasks, despite not being explicitly trained for this purpose. 
Thus, traditional performance metrics, such as the number of solved tasks, are insufficient for isolating the contributions of the decomposition model from those of the synthesis model.
To more effectively attribute performance gains to each component, our work proposes using \ac{regesm} as an ablation method.

Hybrid approaches that integrate transductive and inductive strategies have shown promise in improving program synthesis~\citep{li2024combining}. 
ExeDec exemplifies such a hybrid model, where the Subgoal Model represents the transductive component and the Synthesizer Model embodies the inductive aspect, working together to enable compositional generalization. 
Our results show that both \ac{regesm} and ExeDec successfully solve a diverse range of tasks, suggesting potential benefits of framing ExeDec as a hybrid approach.
However, identifying the specific conditions under which this combination is most effective remains an open question for future research.





\section*{Acknowledgments}
Janis Zenkner is supported by the German Federal Ministry for Digital and Transport (BMDV)
and the German Federal Ministry for Economic
Affairs and Climate Action (BMWK). 

\bibliography{iclr2025_conference}
\bibliographystyle{iclr2025_conference}

\appendix
\newpage
\begin{appendices}
\section{Further ExeDec Examples}
\label{app:examples}
\begin{figure}[h!]
        \begin{subfigure}[b]{\linewidth}
            \begin{minipage}{\textwidth}
            Task specification: {\color{OliveGreen}{$\{x0 = [42, -48]$ $\rightarrow$ $y = [42, 42]$, $x0 = [-35, -21]$ $\rightarrow$ $y = [-35, -21]$, $x0 = [39, 32]$ $\rightarrow$ $y = [39, 39]\}$}}\\
            Ground Truth: \color{BlueViolet}{\texttt{y = Scanl1 (max) x0}}\\
            \end{minipage}
            \begin{minipage}{0.8\textwidth}
                Step 1:
                \begin{itemize}
                    \item Predicted Subgoals:  {\color{OliveGreen}{$[-48, 42], [-35, -21], [32, 29]$}}
                    \item Subprogram: {\color{BlueViolet}{\texttt{x1 = Sort x0}}}
                    \item Update: {\color{OliveGreen}{$x1 = [-48, -42]$, $x1 = [-35, -21]$, $x1 = [32, 39]$}}
                \end{itemize}
            \end{minipage}
            \begin{minipage}{0.8\textwidth}
                Step 2:
                \begin{itemize}
                    \item Predicted Subgoal:  {\color{OliveGreen}{$[48, 42], [-35, -21], [39, 39]$}}
                    \item Subprogram: {\color{BlueViolet}{\texttt{x2 = Zip (max) x0 x1}}}
                    \item Execution:  {\color{OliveGreen}{$x2 = [42, 42]$, $x2 = [-35, -21]$, $x2 = [39, 39]$}}
                \end{itemize}
            \end{minipage}
        \caption{Task 1}
        \vspace{0.25cm}
    \end{subfigure}
    \begin{subfigure}[b]{\linewidth}
        \begin{minipage}{\textwidth}
        Task specification: {\color{OliveGreen}{$\{x0 = 1 | x1 = [-2, -25, 1]$ $\rightarrow$ $y = [-2, -2, 1]$, $x0 = 5 | x1 = -4$ $\rightarrow$ $y = -4$, $x0 = 2 | x1 = [-28, -15]$ $\rightarrow$ $y = [-28, -15]\}$}}\\
        Ground Truth: \color{BlueViolet}{\texttt{y = Scanl1 (max) x0}}\\
        \end{minipage}
        \begin{minipage}{0.8\textwidth}
            Step 1:
            \begin{itemize}
                \item Predicted Subgoals:  {\color{OliveGreen}{$[-25, -2, 1]$, $-4$, $[-28, -19]$}}
                \item Subprogram: {\color{BlueViolet}{\texttt{x2 = Sort x1}}}
                \item Update: {\color{OliveGreen}{$x2 = [-25, -2, 1]$, $x2 = -4$, $x2 = [-28, -19]$}}
            \end{itemize}
        \end{minipage}
        \begin{minipage}{0.8\textwidth}
            Step 2:
            \begin{itemize}
                \item Predicted Subgoal:  {\color{OliveGreen}{$[-25, -23, -24]$, $-4$, $[-28, -13]$}}
                \item Subprogram: {\color{BlueViolet}{\texttt{x3 = Scanl1 (-) x2}}}
                \item Execution:  {\color{OliveGreen}{$x3 = [-25, -23, -24]$, $x2 = -4$, $x2 = [-28, -13]$}}
            \end{itemize}
        \end{minipage}
        \begin{minipage}{0.8\textwidth}
            Step 3:
            \begin{itemize}
                \item Predicted Subgoal:  {\color{OliveGreen}{$[-25, -2, 22]$, $-4$, $[-28, -15]$}}
                \item Subprogram: {\color{BlueViolet}{\texttt{x4 = Scanl1 (-) x3}}}
                \item Execution:  {\color{OliveGreen}{$x4 = [-25, -2, 22]$, $x2 = -4$, $x2 = [-28, -15]$}}
            \end{itemize}
        \end{minipage}
        \begin{minipage}{0.8\textwidth}
            Step 4:
            \begin{itemize}
                \item Predicted Subgoal:  {\color{OliveGreen}{$[-25, -25, 1]$, $-4$, $[-28, -15]$}}
                \item Subprogram: {\color{BlueViolet}{\texttt{x5 = Zip (min) x1 x4}}}
                \item Execution:  {\color{OliveGreen}{$x5 = [-25, -25, 1]$, $x2 = -4$, $x2 = [-28, -15]$}}
            \end{itemize}
        \end{minipage}
        \begin{minipage}{0.8\textwidth}
            Step 5:
            \begin{itemize}
                \item Predicted Subgoal:  {\color{OliveGreen}{$[-2, -25, 1]$, $-4$, $[-28, -15]$}}
                \item Subprogram: {\color{BlueViolet}{\texttt{x6 = Zip (max) x1 x5}}}
                \item Execution:  {\color{OliveGreen}{$x6 = [-2, -25, 1]$, $x6 = -4$, $x6 = [-28, -15]$}}
            \end{itemize}
        \end{minipage}
        \begin{minipage}{0.8\textwidth}
            Step 6:
            \begin{itemize}
                \item Predicted Subgoal:  {\color{OliveGreen}{$[-2, -2, 1]$, $-4$, $[-28, -15]$}}
                \item Subprogram: {\color{BlueViolet}{\texttt{x7 = Zip (max) x2 x6}}}
                \item Execution:  {\color{OliveGreen}{$x7 = [-2, -2, 1]$, $x7 = -4$, $x7 = [-28, -15]$}}
            \end{itemize}
        \end{minipage}
        \caption{Task 2}
    \end{subfigure}
    \caption{Examples from the Deepcoder domain displaying the struggles of misleading subtasks. Both tasks can be solved with a single-step function that computes the cumulative maximum from the input list.}
    \label{fig:app_example}
\end{figure}

\section{Robustfill and Deepcoder DSLs}
\label{app:dsls}
Both \acp{dsl} were taken from the original ExeDec study~\citep{shi2023exedec}.
\begin{figure}[h!]
    \centering
    \[
    \begin{array}{rl}
        \text{Program} \; P := & \texttt{Concat}(e_1, e_2, \ldots) \\
        \text{Expression} \; e := & s \;|\; m \;|\; o \;|\; \texttt{ConstStr}(c) \\
        \text{Compose} \; o := & m_1(m_2) \;|\; m(s) \\
        \text{Substring} \; s := & \texttt{SubStr}(k_1, k_2) \;|\; \texttt{GetSpan}(r_1, i_1, b_1, r_2, i_2, b_2) \\
                                    & \;|\; \texttt{GetUpto}(r, i) \;|\; \texttt{GetFrom}(r, i) \;|\; \texttt{GetToken}(r, i) \\
        \text{Modification} \; m := & \texttt{ToCase}(a) \;|\; \texttt{Replace}(c_1, c_2) \;|\; \texttt{Trim()} \\
                                      & \;|\; \texttt{GetFirst}(r, i) \;|\; \texttt{GetAll}(r) \\
                                      & \;|\; \texttt{Substitute}(r, i, c) \;|\; \texttt{SubstituteAll}(r, c) \\
                                      & \;|\; \texttt{Remove}(r, i) \;|\; \texttt{RemoveAll}(r) \\
        \text{Regex} \; r := & \texttt{NUMBER} \;|\; \texttt{WORD} \;|\; \texttt{ALPHANUM} \;|\; \texttt{ALL\_CAPS} \;|\; \texttt{PROPER\_CASE} \\
                                & \;|\; \texttt{LOWER} \;|\; \texttt{DIGIT} \;|\; \texttt{CHAR} \;|\; \delta \\
        \text{Case} \; a := & \texttt{ALL\_CAPS} \;|\; \texttt{PROPER\_CASE} \;|\; \texttt{LOWER} \\
        \text{Position} \; k := & -100 \;|\; -99 \;|\; \ldots \;|\; -1 \;|\; 0 \;|\; 1 \;|\; 2 \;|\; \ldots \;|\; 100 \\
        \text{Index} \; i := & -5 \;|\; -4 \;|\; \ldots \;|\; -1 \;|\; 1 \;|\; 2 \;|\; \ldots \;|\; 5 \\
        \text{Boundary} \; b := & \texttt{START} \;|\; \texttt{END} \\
        \text{Character} \; c := & A \;|\; \ldots \;|\; Z \;|\; a \;|\; \ldots \;|\; z \;|\; 0 \;|\; \ldots \;|\; 9 \;|\; \delta \\
        \text{Delimiter} \; \delta := & \texttt{\& , . ?  ! @ ()  [] \% \# \$ " ´} \\
    \end{array}
    \]
    \caption{String manipulation functions that can be used to solve tasks from the Robustfill domain.}
    \label{fig:dsl_rf}
\end{figure}

\begin{figure}[h]
    \centering 
    \[
    \begin{array}{rl}
        \text{Program} \; P := & i_1; \; i_2; \; \ldots; \; a_1; \; a_2; \; \ldots \\[8pt]
        \text{Initialization} \; i := & v \gets \texttt{INPUT} \\[8pt]
        \text{Assignment} \; a := & v \gets f \;|\; v \gets h \\[8pt]
        \text{First-Order Operation} \; f := & \texttt{Head}(l) \;|\; \texttt{Last}(l) \;|\; \texttt{Access}(n, l) \;|\; \texttt{Minimum}(l) \;|\; \texttt{Maximum}(l) \\
                                              & \;|\; \texttt{Sum}(l) \;|\; \texttt{Take}(n, l) \;|\; \texttt{Drop}(n, l) \;|\; \texttt{Reverse}(l) \;|\; \texttt{Sort}(l) \\[8pt]
        \text{Higher-Order Operation} \; h := & \texttt{Map}(\lambda, l) \;|\; \texttt{Filter}(\beta, l) \;|\; \texttt{Count}(\beta, l) \;|\; \texttt{Zip}(\Sigma, l, l) \\
                                                 & \;|\; \texttt{Scanl1}(\Sigma, l) \\[8pt]
        \text{int} \to \text{int} \; \text{Lambda} \; \lambda := & (+1) \;|\; (-1) \;|\; (*2) \;|\; (/2) \;|\; (*(-1)) \;|\; (**2) \;|\; (*3) \;|\; (/3) \;|\; (*4) \;|\; (/4) \\[8pt]
        \text{int} \to \text{bool} \; \text{Lambda} \; \beta := & (> 0) \;|\; (< 0) \;|\; (\%2 == 0) \;|\; (\%2 == 1) \\[8pt]
        (\text{int}, \text{int}) \to \text{int} \; \text{Lambda} \; \Sigma := & (+) \;|\; (-) \;|\; (*) \;|\; (\texttt{min}) \;|\; (\texttt{max}) \\[8pt]
        \text{Integer Variable} \; n := & v \\[8pt]
        \text{List Variable} \; l := & v \\[8pt]
        \text{Variable Name} \; v := & x_1 \;|\; x_2 \;|\; \ldots \\
    \end{array}
    \]
    \caption{First and Higher-Order Functions contained in the \ac{dsl} for the Deepcoder domain.}
    \label{fig:dsl_dc}
\end{figure}

\section{Benchmark Creation}
\label{app:benchmarks}
\cite{shi2023exedec} define tasks based on compositional generalization categories by structuring programs according to their length and the concepts they employ.

\subsection{Robustfill}
In the Robustfill domain, each program is composed of concatenated subprograms, and program length is determined by the number of these subprograms.
\begin{itemize}
    \item Length Generalization: Training tasks consist of programs with lengths ranging from 1 to 6, while testing tasks feature longer programs with lengths between 7 and 10.
    \item Compose Different Concepts: The operations are categorized into two concepts: substring-related operations form the “substring concept,” while modification operations and constant strings belong to the “non-substring concept” (excluding the Compose operation). Both training and testing programs have lengths between 2 and 6.
    \item Switch Concept Order: Training tasks enforce a structure where the first half of the subprograms belong to the substring concept, and the latter half to the non-substring concept. In testing tasks, this order is reversed. Program lengths remain between 2 and 6.
    \item Compose-New-Operation: A quarter of the training tasks are single-subprogram programs that exclusively contain the \texttt{Compose} operation. The remaining training tasks consist of programs of length 2-6 without Compose. Testing tasks include length 2-6 programs that incorporate the \texttt{Compose} operation.
    \item Add Operation Functionality: Training tasks consist of programs of length 1-6 where substring operations are never used within the \texttt{Compose} operation. Testing tasks maintain the same length range but include cases where substring operations appear within \texttt{Compose}.
\end{itemize}

\subsection{Deepcoder}
In the Deepcoder domain, programs are line-by-line and task length is defined as the number of non-input lines in the program.
\begin{itemize}
    \item 	Length Generalization: Training tasks consist of programs with lengths between 1 and 4, while testing tasks feature programs of length 5.
    \item Compose Different Concepts: Operations are divided into two categories: the first concept includes all first-order operations and the \texttt{Map} operation, while the second concept consists of all higher-order operations. Both training and testing programs have lengths between 1 and 4.
    \item Switch Concept Order: Training tasks follow a structure where the first half of the subprograms belong to the first-order operations and \texttt{Map} concept, while the latter half use higher-order operations. In testing tasks, this order is reversed. Program lengths remain between 1 and 4.
    \item Compose New Operation: A quarter of the training tasks consist of length 1 programs that exclusively contain the \texttt{Scanl1} operation. The remaining training tasks include length 2-4 programs that do not use \texttt{Scanl1}. Testing tasks include length 2-4 programs that incorporate the \texttt{Scanl1} operation.
    \item Add Operation Functionality: Training tasks consist of programs of length 1-4, where \texttt{Scanl1} is only applied with the lambda functions \texttt{(-)} and \texttt{(min)}. In testing tasks, \texttt{Scanl1} is used with additional lambda functions \texttt{(+)}, \texttt{(*)}, and \texttt{(max)}.
\end{itemize}
\section{Compositional Generalization Analysis}
\label{app:results}
In the Robustfill domain, the majority of tasks is concentrated along $x=100\%$, representing tasks where subtasks are predicted as intended by the ground truth but the implementation differs, showcasing the Subgoal Model’s accuracy in learning decompositions.
The densities decrease as you move towards the bottom-left corner, indicating fewer cases that deviate semantically and syntactically from the ground truth.
Yet, these tasks are solved using - at least partially - alternative decompositions.
Interestingly, the densities do not differ clearly between the compositional generalization tasks, even though the performance varies significantly.
\begin{figure}[h]
    \centering
    \begin{subfigure}[b]{0.29\linewidth}
        \centering
        \includegraphics[width=\linewidth]{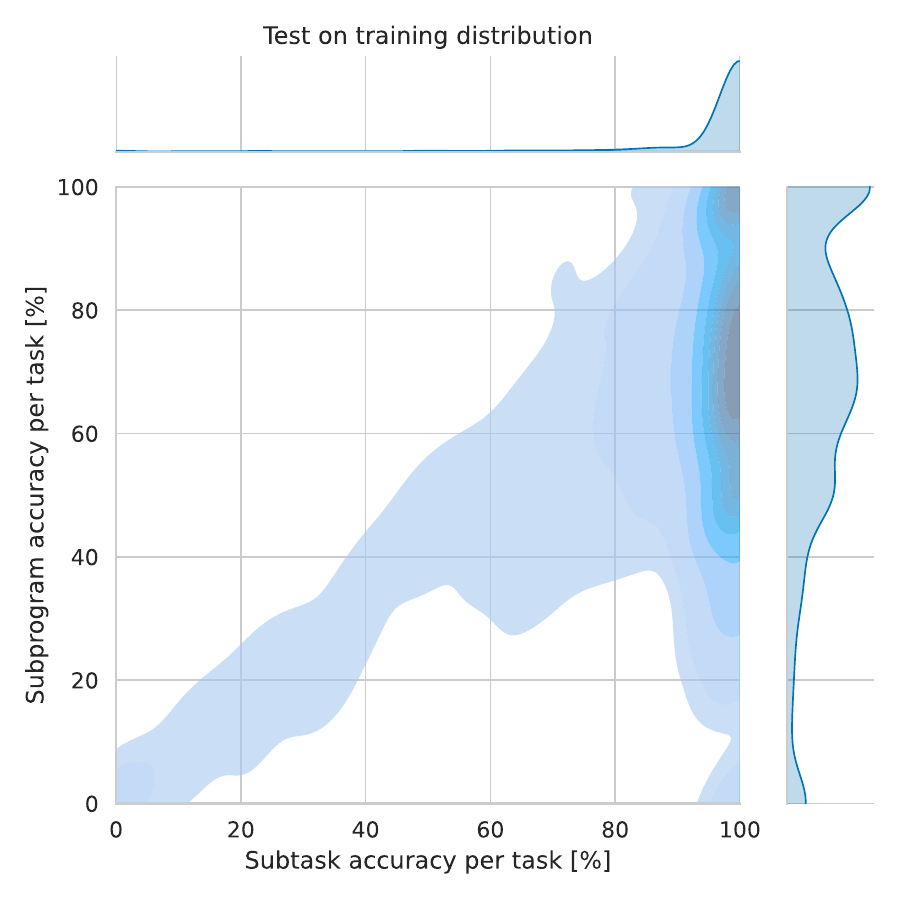}
        \caption{Test on training distribution}
    \end{subfigure}
    \begin{subfigure}[b]{0.29\linewidth}
        \centering
        \includegraphics[width=\linewidth]{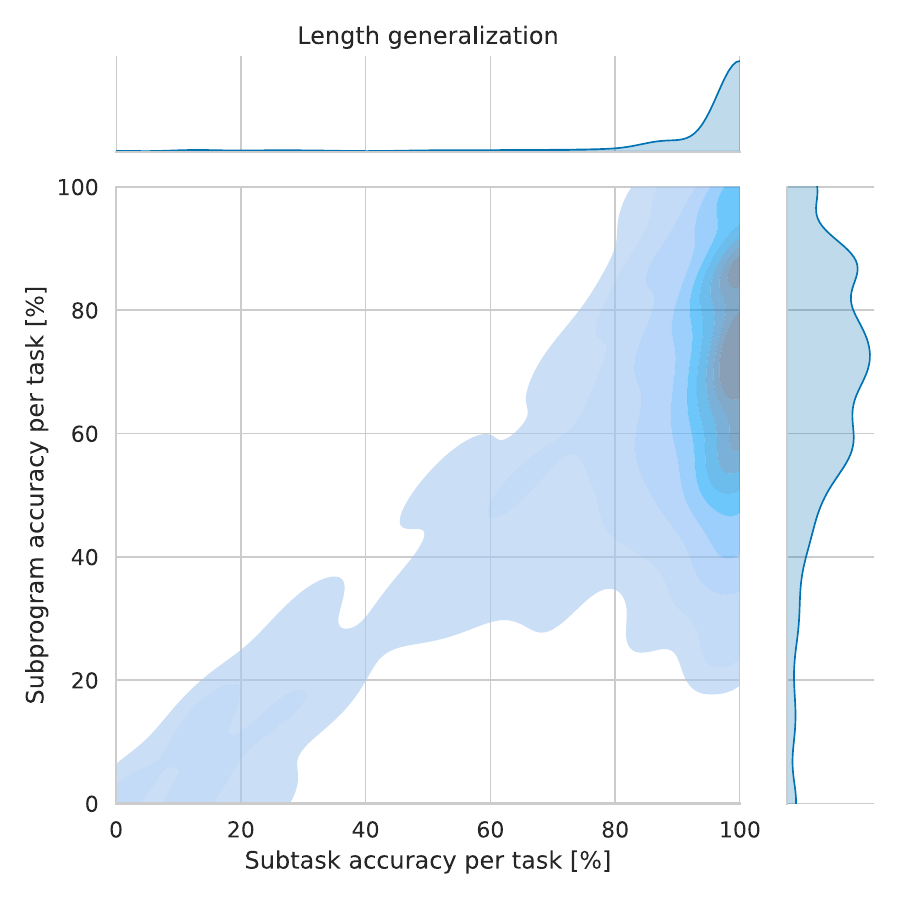}
        \caption{Length generalization}
    \end{subfigure}
    \begin{subfigure}[b]{0.29\linewidth}
        \centering
        \includegraphics[width=\linewidth]{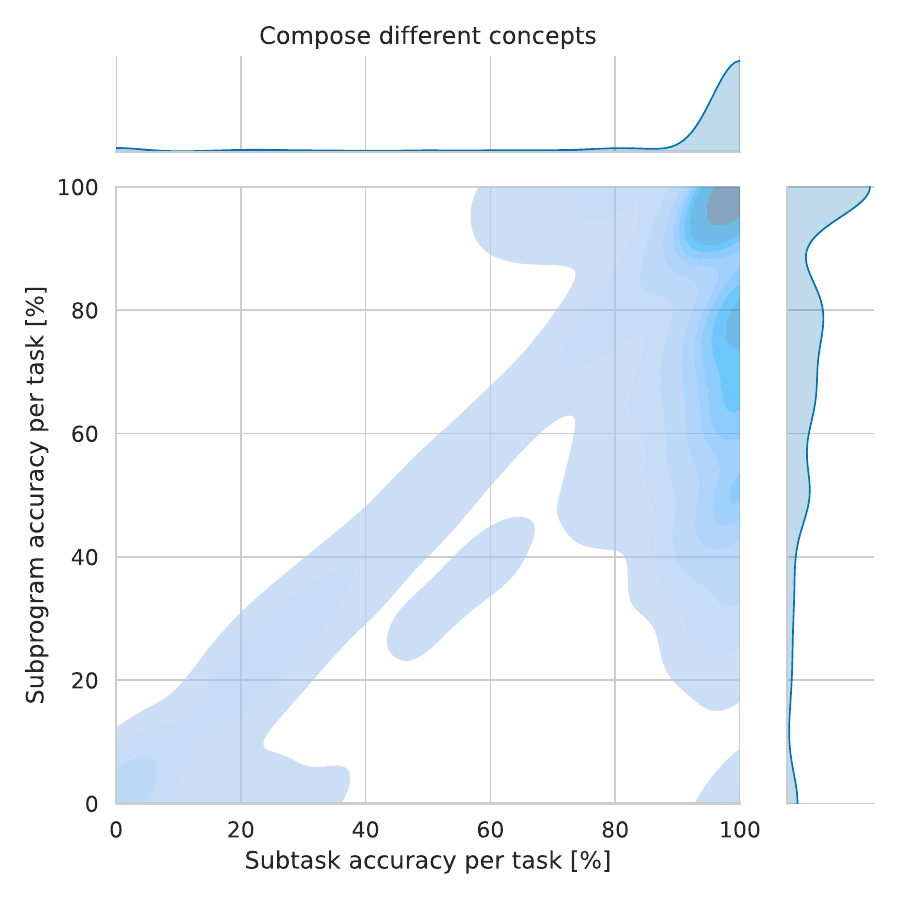}
        \caption{Compose different concepts}
    \end{subfigure}
    \begin{subfigure}[b]{0.29\linewidth}
        \centering
        \includegraphics[width=\linewidth]{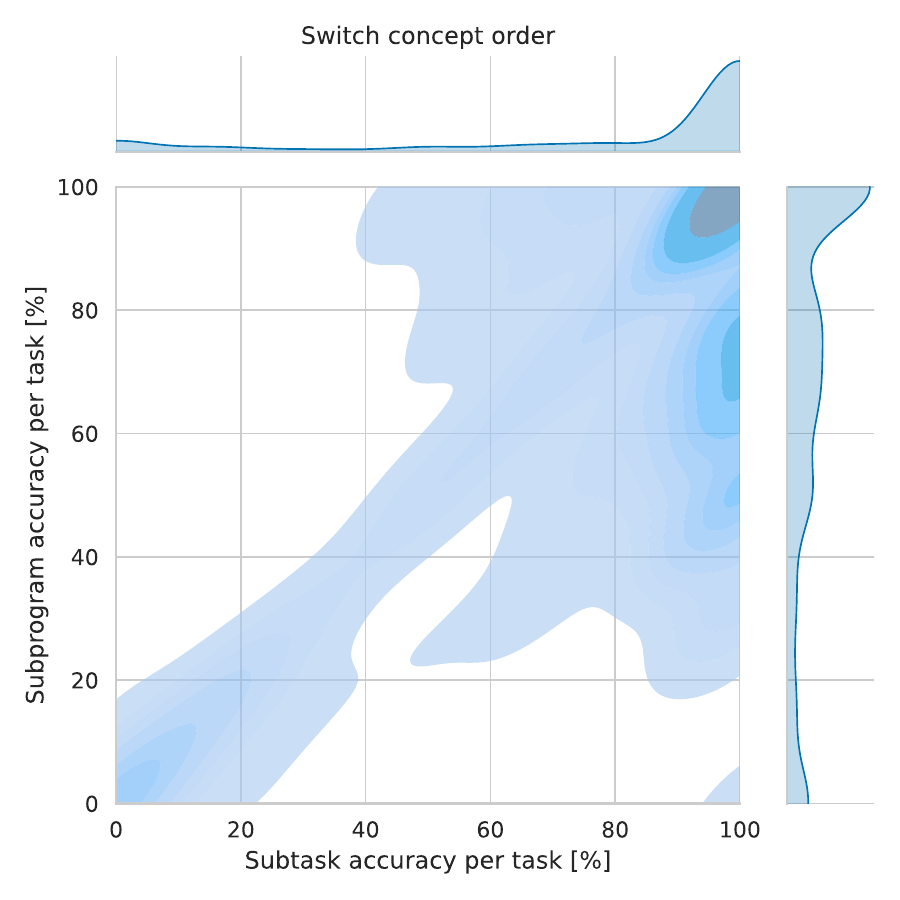}
        \caption{Switch concept order}
    \end{subfigure}
    \begin{subfigure}[b]{0.29\linewidth}
        \centering
        \includegraphics[width=\linewidth]{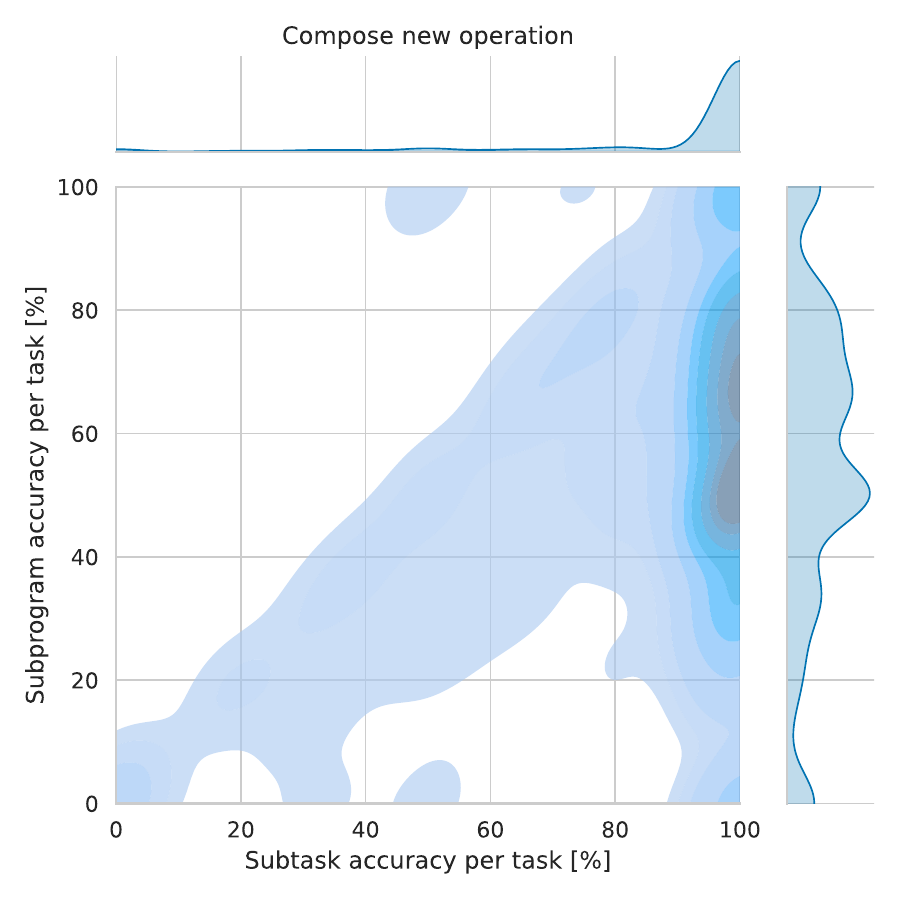}
        \caption{Compose new operation}
    \end{subfigure}
    \begin{subfigure}[b]{0.29\linewidth}
        \centering
        \includegraphics[width=\linewidth]{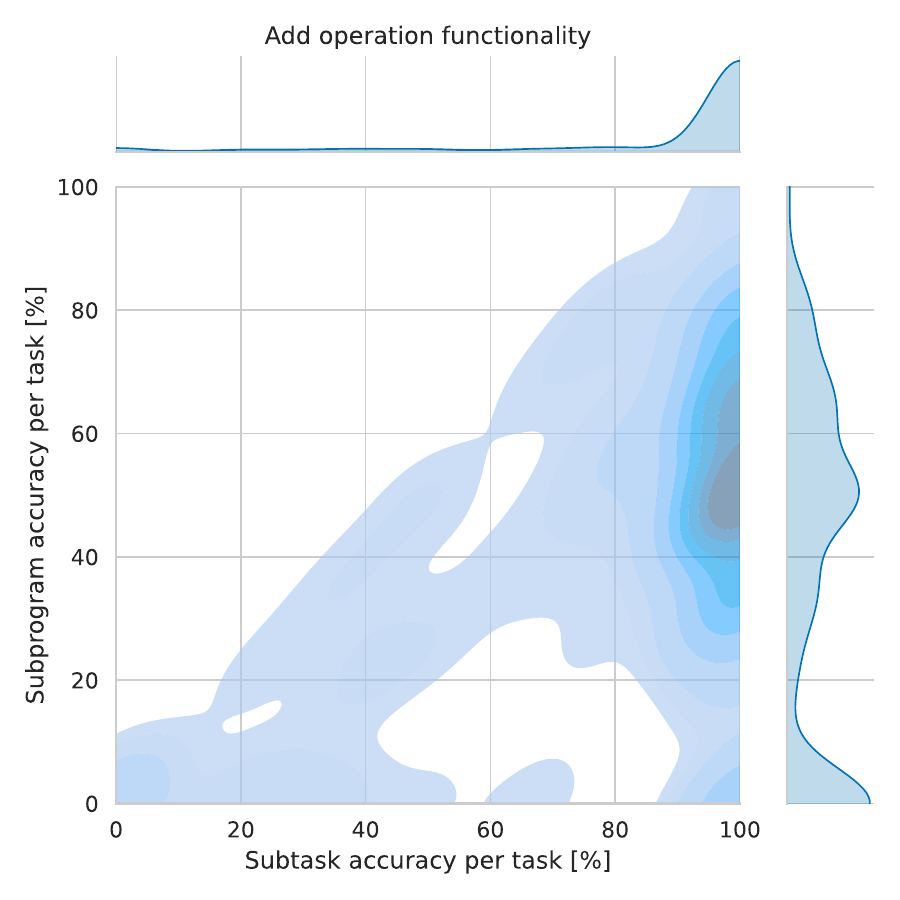}
        \caption{Add operation functionality}
    \end{subfigure}
    \caption{Solved tasks clustered across subtask and subprogram accuracy.}
    \label{fig:density_all_concepts_robustfill}
\end{figure}

\begin{figure}[h]
    \centering
    \begin{subfigure}[b]{0.29\linewidth}
        \centering
        \includegraphics[width=\linewidth]{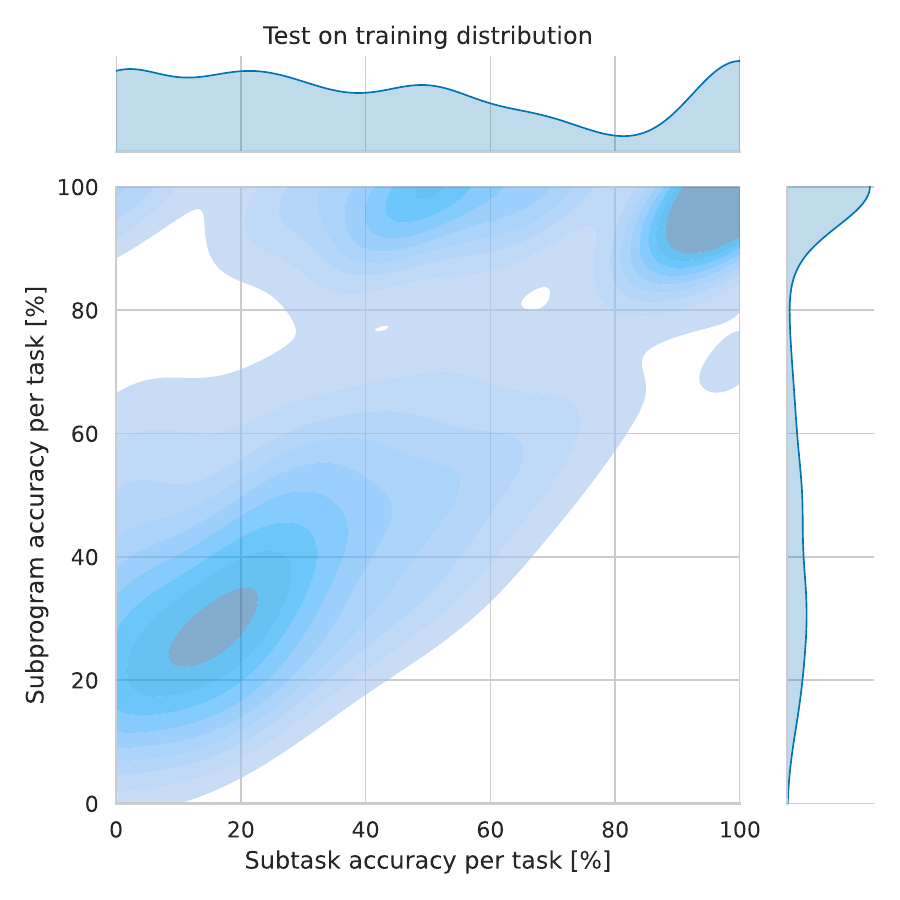}
        \caption{Test on training distribution}
    \end{subfigure}
    \begin{subfigure}[b]{0.29\linewidth}
        \centering
        \includegraphics[width=\linewidth]{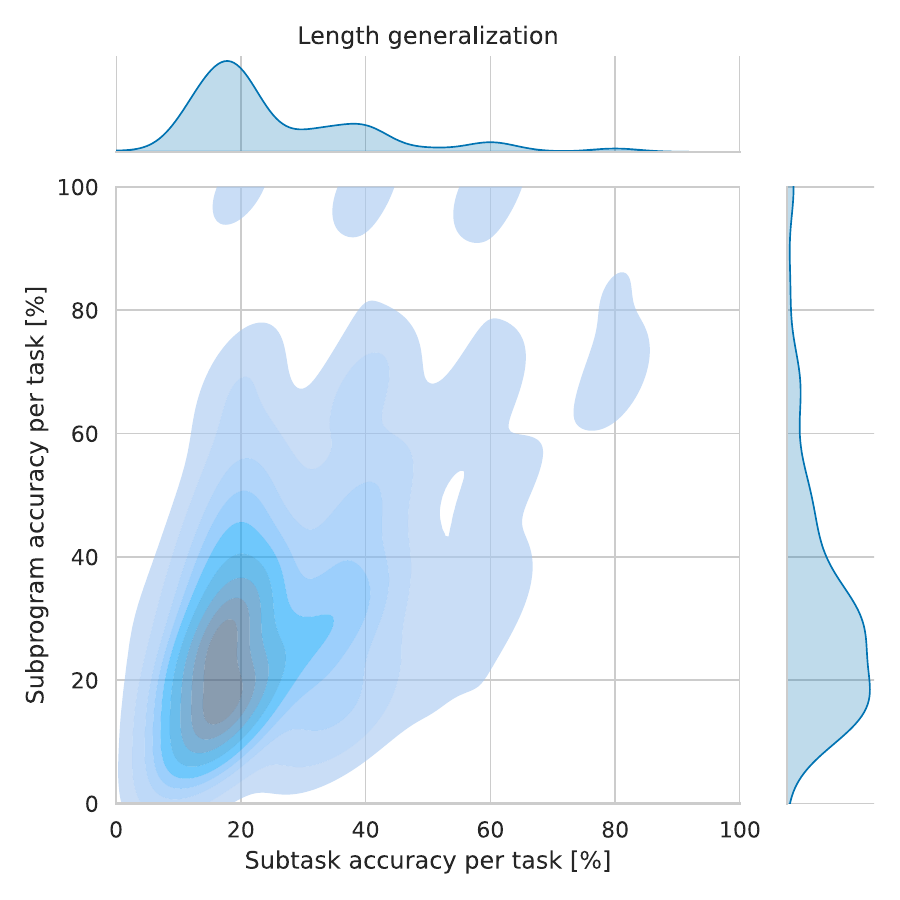}
        \caption{Length generalization}
    \end{subfigure}
    \begin{subfigure}[b]{0.29\linewidth}
        \centering
        \includegraphics[width=\linewidth]{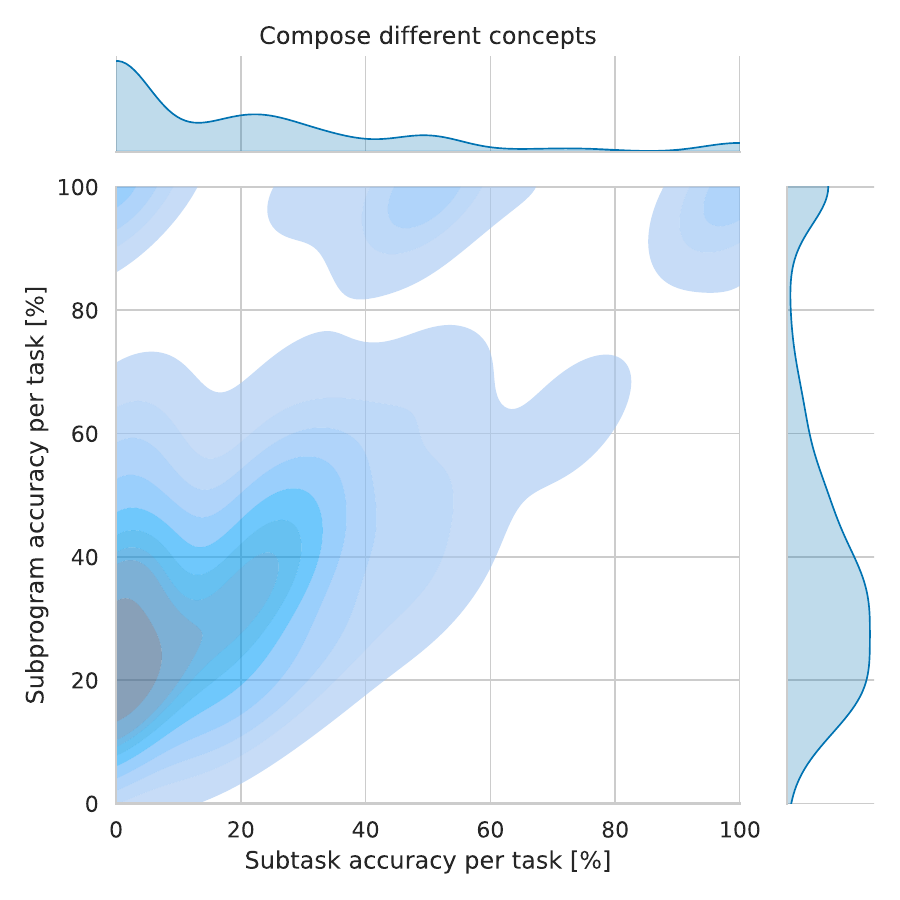}
        \caption{Compose different concepts}
    \end{subfigure}
    \begin{subfigure}[b]{0.29\linewidth}
        \centering
        \includegraphics[width=\linewidth]{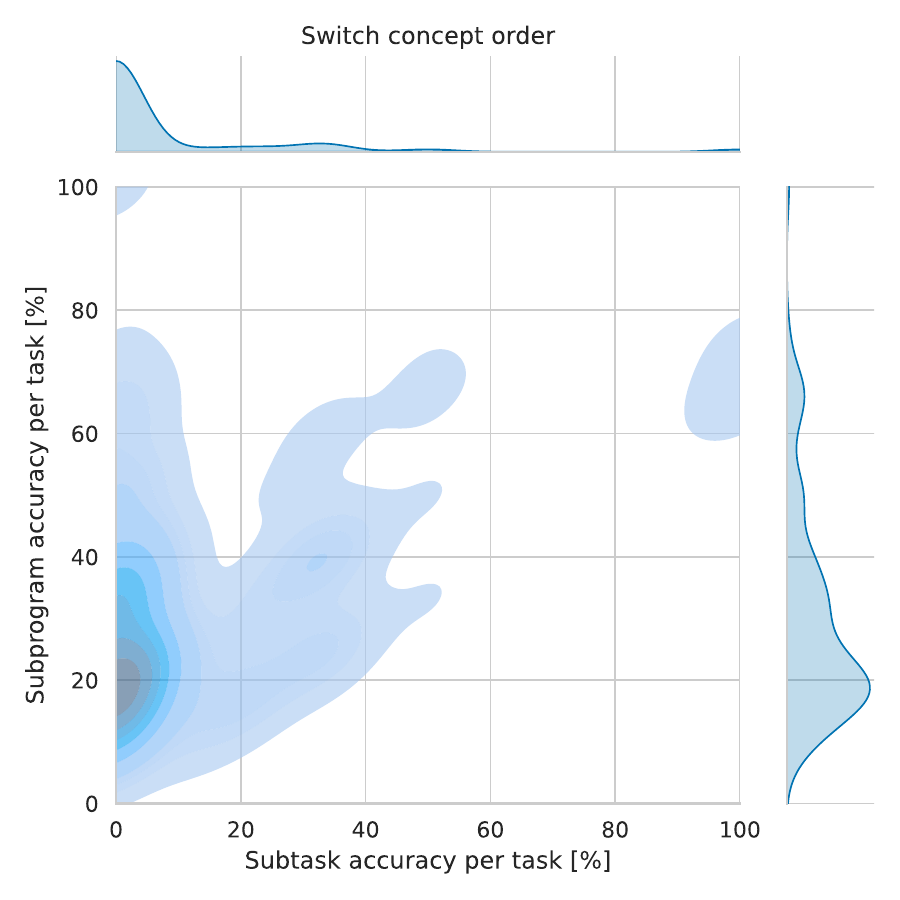}
        \caption{Switch concept order}
    \end{subfigure}
    \begin{subfigure}[b]{0.29\linewidth}
        \centering
        \includegraphics[width=\linewidth]{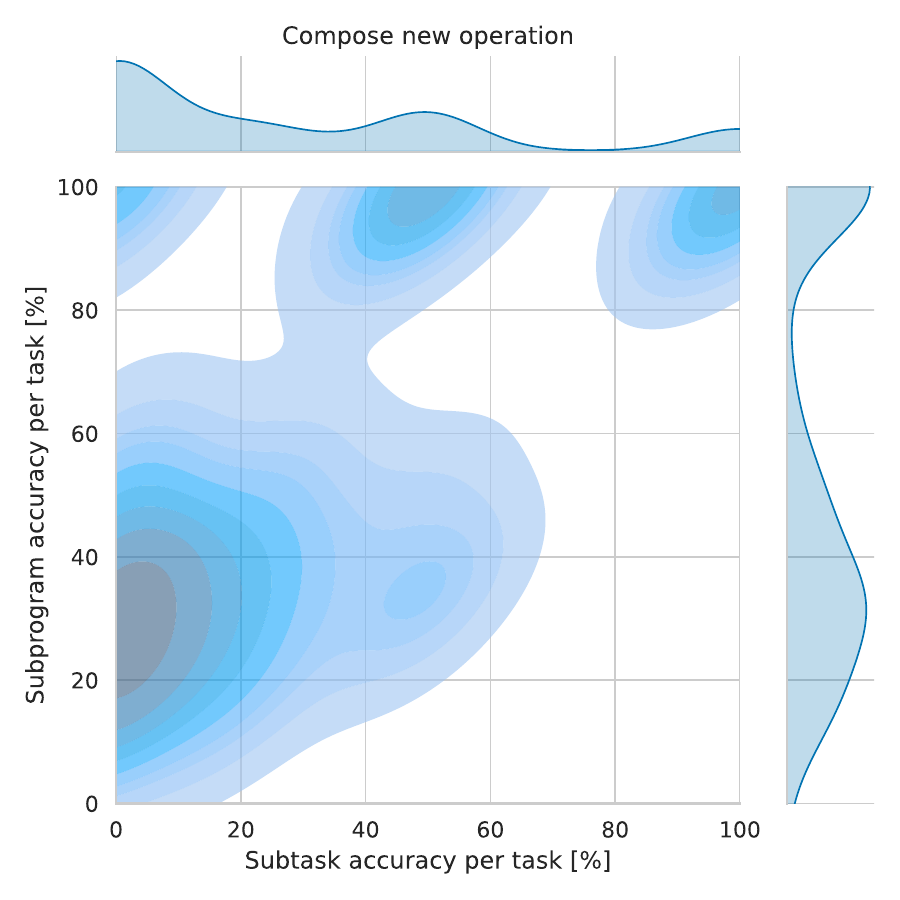}
        \caption{Compose new operation}
    \end{subfigure}
    \begin{subfigure}[b]{0.29\linewidth}
        \centering
        \includegraphics[width=\linewidth]{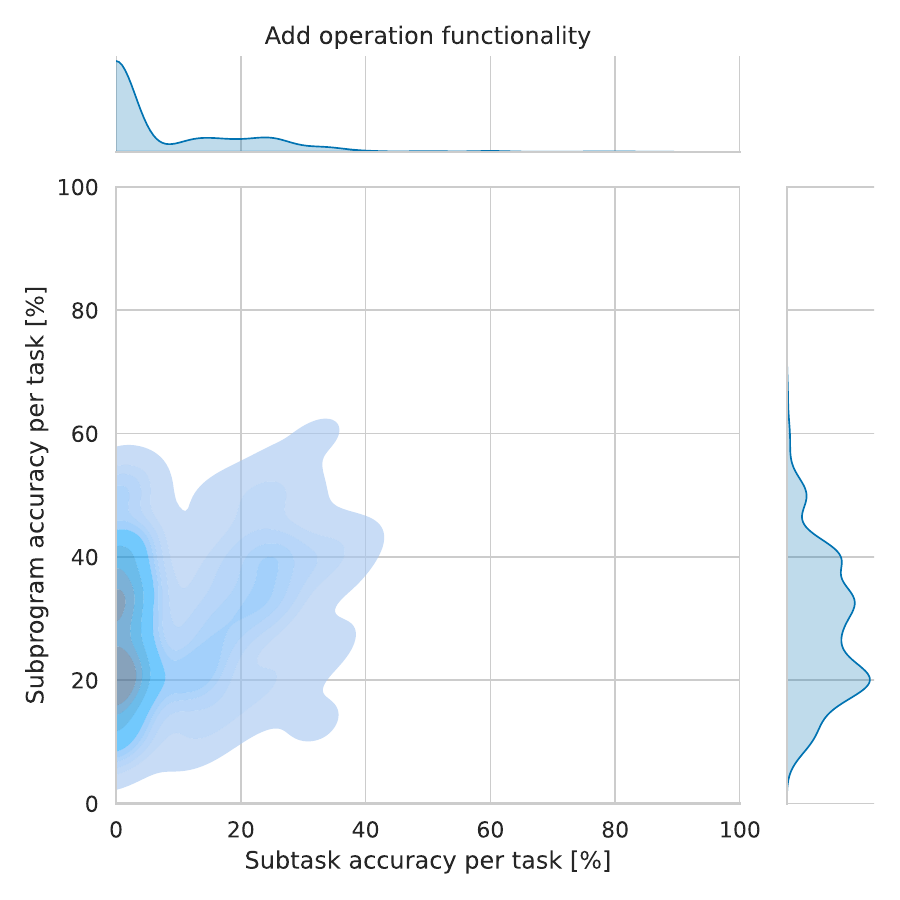}
        \caption{Add operation functionality}
    \end{subfigure}
    \caption{Solved tasks clustered across subtask and subprogram accuracy.}
    \label{fig:density_all_concepts_deepcoder}
\end{figure}
In the Deepcoder domain, four distinct clusters can be observed. While small clusters exist in the top-right corner, indicating tasks solved as intended by the ground truth, most tasks are solved by using alternative decompositions and implementations (bottom-left corner).
Again, no clear differences between the compositional generalization tasks exist.
Yet, the subgoal accuracy distribution is more uniform when tested on the training distribution.
This indicates that the Subgoal Model can learn and apply decomposition patterns in this task category more easily.

\begin{figure}[h]
    \centering
    \begin{subfigure}[b]{0.7\linewidth}
        \centering
        \includegraphics[width=\linewidth]{plots/num_decompositions_barplot_deepcoder_ExeDec.pdf}
        \caption{Deepcoder Domain}
        \label{fig:num_decomps_deepcoder}
    \end{subfigure}
    \begin{subfigure}[b]{0.7\linewidth}
        \centering
        \includegraphics[width=\linewidth]{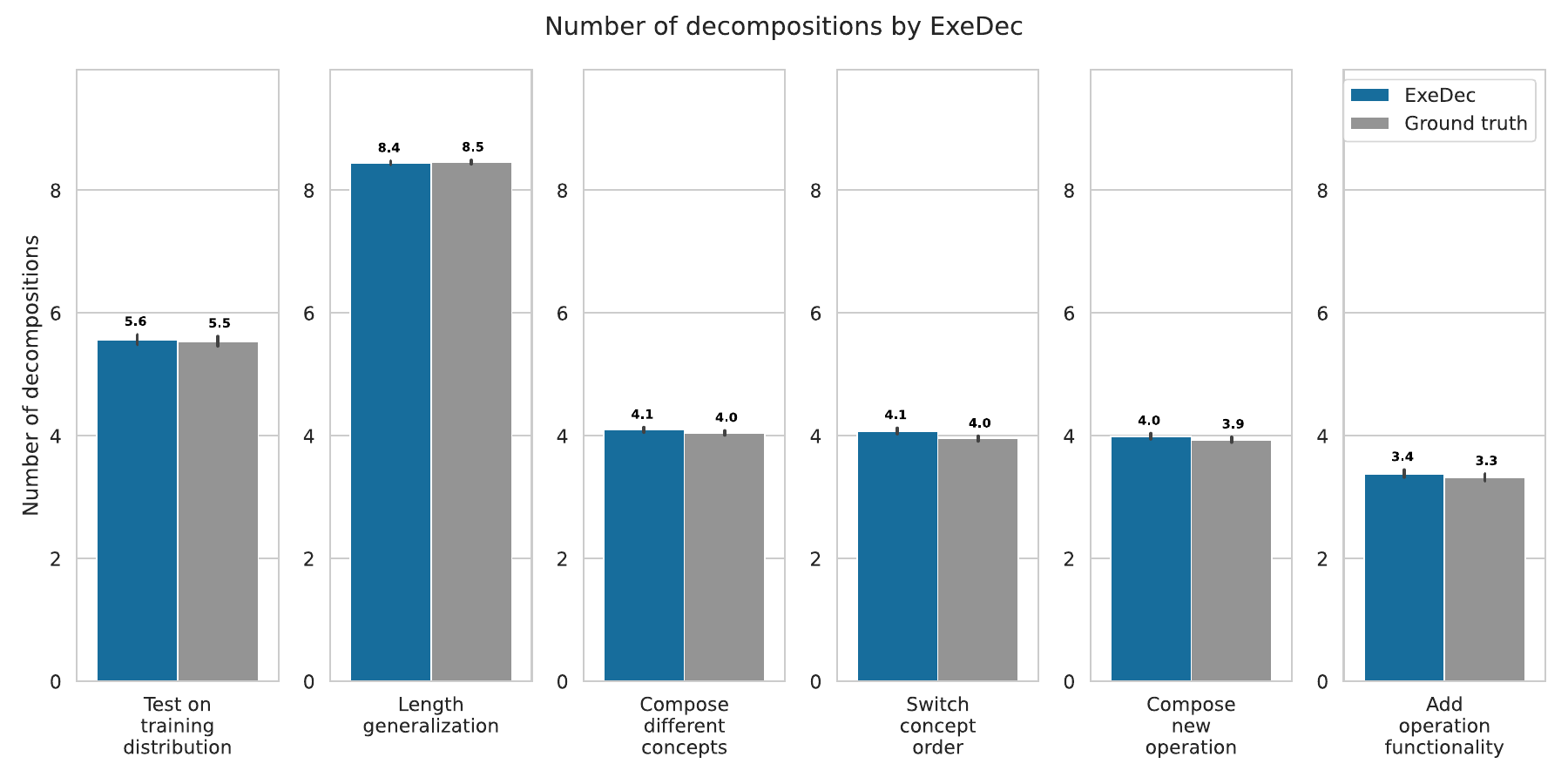}
        \caption{Robustfill Domain}
        \label{fig:num_decomps_robustfill}
    \end{subfigure}
    \caption{Comparison of the number of decompositions using ExeDec.}
    \label{fig:app_num_decomp}
\end{figure}

The differences between the Robustfill and Deepcoder domains are evident in the decomposition process.
This becomes again clear when comparing the number of decompositions to the ground truth.
In the Robustfill domain, the number of predicted decompositions aligns closely with the ground truth, where the predicted solutions almost exactly match the number of decompositions required by the target solution (Figure~\ref{fig:num_decomps_robustfill}). 
This suggests that the solutions closely match the target program, likely due to the relatively straightforward nature of decomposition learning in the Robustfill domain.
In contrast, ExeDec’s Subgoal Model in the Deepcoder domain decomposes tasks into significantly more subtasks than specified by the ground truth (Figure~\ref{fig:num_decomps_deepcoder_exedec}).
This leads to more frequent invocations of the Synthesizer Model, which appears to be a key driver of the observed performance improvements. 
The increased number of decompositions reflects the greater challenge of learning decomposition in the Deepcoder domain, where tasks often require a more nuanced understanding and finer granularity of subgoal generation.

Furthermore, the simpler structure of the Robustfill domain, where intermediate states remain constant, contrasts with Deepcoder’s dynamic intermediate states and complex task structures. 
This dynamic nature better mirrors how humans approach coding tasks.
However, the Subgoal Model performs significantly worse in the Deepcoder domain, highlighting challenges in capturing this human-like reasoning.
Due to the ambiguous performance of the Subgoal Model in the Deepcoder domain, we focused on this domain in subsequent experiments to better understand the factors influencing its behavior and to explore potential enhancements.
\end{appendices}

\end{document}